\documentclass[aps, prx, reprint, superscriptaddress, 10pt]{revtex4-2}
\usepackage{amsmath,amsfonts,amsthm,amssymb}
\usepackage{graphicx}
\usepackage[caption=false]{subfig}
\usepackage{placeins}
\usepackage{mathrsfs}
\usepackage{etoolbox}

\usepackage[colorlinks,linkcolor=blue,citecolor=blue,urlcolor=blue]{hyperref}

\renewcommand{\vec}[1]{\boldsymbol{#1}}

\begin{document}

\title{Kinetic Theory for the Shear Viscosity of Dense Binary Dipolar Fluid Mixtures}
\author{Christopher Devik Fjeldstad}
\affiliation{Department of Mechanical and Industrial Engineering, Norwegian University of Science and Technology, NO-7491 Trondheim, Norway}
\author{Roberto E. Troncoso}
\affiliation{Instituto de Alta Investigaci\'on, Universidad de Tarapac\'a, Casilla 7D, Arica, Chile}
\author{Astrid S. de Wijn}
\affiliation{Department of Mechanical and Industrial Engineering, Norwegian University of Science and Technology, NO-7491 Trondheim, Norway}

\begin{abstract}
We construct a kinetic theory for the shear viscosity of dense binary fluid mixtures of strongly-interacting dipolar hard spheres.
We derive an expression for the pairwise correlations in the binary mixtures that is  accurate up to packing fractions around 0.35.
The approach is based on Enskog-Thorne theory, and inspired by the theory for dense pure fluids developed by Pousaneh and de Wijn.
It relies on effective coupling parameters obtained from the pure fluids combined with mixing rules and a heuristic expression for the collision integral.
We compare our results to viscosities obtained numerically from molecular-dynamics simulations of dipolar hard-sphere fluids.
Our expression for the shear viscosity of the binary mixtures captures the density and composition dependent behavior of the binary dipolar fluids up to packing fraction of $\xi \lesssim 0.3$ without any mixture-derived fit parameters.
\end{abstract}

\maketitle

\section{Introduction}

Quantitative predictions of transport properties of fluids, such as the shear viscosity, present a non-trivial problem in non-equilibrium statistical mechanics.
While transport properties are of crucial importance in many applications, engineers are often forced to use phenomenological ad-hoc approaches involving large numbers of adjustable parameters.
Most of the ad-hoc approaches are based on kinetic theory of gases and liquids  given by Boltzmann's transport equation. They combine Chapman's approximate solutions with Enskog's heuristic extension to dense fluids~\cite{Chapman1952}. Although initially limited to pure fluids, Enskog-Thorne theory~\cite{Chapman1952} extended the Enskog approach to include mixtures.
The approach of Enskog is however limited in its applicability due to the extreme short-range nature of the interactions it can describe.
This precludes fluids with electrostatic interactions, including those with dipoles (e.g.\ CO, water, ethanol, and most organic solvents) and quadrupoles (e.g.\ CO$_2$), many of which are of considerable practical interest, especially as part of mixtures.

At low densities, there are several methods for predicting the composition dependent shear viscosity of gaseous mixtures, for example the one proposed by Wilke~\cite{Wilke1950}, which has been successfully applied to polar gas mixtures~\cite{Mathur}.
However, at intermediate and high densities, the accuracy of the Wilke method is limited.
Methods more suitable for such densities, by Di Pippo \textit{et al.}~\cite{di_pippo}, Kestin \textit{et al.}~\cite{Kestin1981}, and Vesovic \textit{et al.}~\cite{Vesovic1989} have demonstrated that Enskog-Thorne theory can successfully be applied to real dense but non-polar mixtures.

Substantial effort has been made to extend Chapman-Enskog theory in ways that account for the presence of non-rigid interactions. 
Of special interest in the present work is the shear viscosity of fluids that contain long-range dipolar interactions. Previously, Pousaneh and de Wijn~\cite{Pousaneh2} have developed a kinetic theory for the shear viscosity of pure dipolar hard-sphere (DHS) fluids. Their approach relies on an expression for the hard-core collision rate that takes into account the impact of long-range dipolar interactions on the structure of the DHS fluid.

In the present work, we extend the work by Pousaneh and de Wijn to binary mixtures of particles using Enskog-Thorne theory. This requires the development of sufficiently accurate expressions for the pairwise correlations of DHS particles in such binary mixtures. To this end, we derive an expression for the partial pair distribution function based on the re-summation of the cluster expansion coefficients derived for binary DHS fluids by Novak~\textit{et al.}~\cite{Novak2013}. The approach was inspired by a similar approach by Elfimova~\textit{et al.} for calculating the free energy of DHS fluids~\cite{Elfimova2012}.
In order to properly test the limits of our proposed expression, we compare the predictions to molecular dynamics simulations.

A description of the microscopic fluid model is given in Sec.~\ref{sec:PhysMod}.
Sec.~\ref{sec:Theory} provides an introduction to the most relevant expressions that our approach builds on.
The new approach is derived in Sec.~\ref{sec:Dipolar_DHS_theory}.
Special interest is given to the partial pair distribution function in Sec.~\ref{sec:PDF_DHS}. 
Simulation details can be found in Sec.~\ref{sec:SimSetUp}. The resulting comparison can be found in Sec.~\ref{sec:results}.

\section{Microscopic Model}
\label{sec:PhysMod}

We consider binary mixtures of dipolar hard spheres (DHS).
The system is characterized by a fixed volume $V$, a temperature $T$, and a number density~$\rho = N/V$, resulting in a pressure~$P$.
For a species $i$, $N_i$ denotes the number of particles of type $i$ in the fluid so that the total number of particles in the binary fluid mixture $N = N_1 + N_2$ and the number fraction $x_i = N_i / N$.
The hard-sphere diameter of species $i$ is denoted $\sigma_i$, the mass by $m_i$ and the dipole moment by $\mu_i$.
The total fraction of volume occupied by the hard cores $\xi$ is given by $\xi = \pi\rho\sum_i x_i\sigma_i^3/6$.
Because dipoles are present, and therefore rotational dynamics are involved, the particles are assigned a moment of inertia $I_i$.

The DHS interaction is given as the sum of a hard-core interaction and interactions of point dipoles located at the centers of the spheres. 
For two interacting particles of species $i$ and $j$, the hard-core collision distance $\sigma_{ij}$ is given by $\sigma_{ij} = (\sigma_i + \sigma_j)/2$.
The interaction energy between two hard spheres is expressed in terms of $\sigma_{ij}$ so that
\begin{equation}
    u^\mathrm{HS}_{ij} = 
    \begin{cases}
        \infty, \quad & r < \sigma_{ij} \\
        0,       \quad & r \geq \sigma_{ij}
    \end{cases}~,
\end{equation}
where $r$ is the inter-particle distance. The point dipole interaction energy is given by 
\begin{equation}
    \label{eq:u_D}
    u^\mathrm{D}_{ij} = \frac{1}{4\pi\epsilon_0}\left[ \frac{\vec{\mu}_i\cdot{\vec \mu}_j}{r^3}-\frac{3(\vec{\mu}_i\cdot \vec{r})(\vec{\mu}_j\cdot \vec{r})}{r^5} \right]~,
\end{equation}
where $\vec{\mu}_i$ and $\vec{\mu}_j$ are the two dipole vectors, $\vec{r}$ is the relative position in space, and $\epsilon_0$ is the permittivity of vacuum. 
For convenience,
we define a dimensionless dipolar coupling constant $\lambda_{ij}$ between two interacting DHS particles as
\begin{align}
    \label{eq:lambda_ij}
    \lambda_{ij}  = \beta\frac{{\mu_i}{\mu_j}}{4\pi\epsilon_0\sigma_{ij}^3}~, 
\end{align}
where $\beta = 1/k_\mathrm{B}T$ where $k_\mathrm{B}$ is Boltzmann's constant.

\section{Theoretical Foundation}
\label{sec:Theory}

Chapman-Enskog theory and its extensions form the basis of our calculations.
In this section, we introduce a number of key concepts and results of Chapman-Enskog theory.
It is based on Boltzmann transport theory to derive hydrodynamic equations for the transport of mass, heat and momentum, the latter of which is related to the shear viscosity $\eta$.
An approximation for the low-density limit of the shear viscosity $\eta^0_i = \lim_{\rho \to 0} \eta$ of a pure fluid is given by the Chapman expression~\cite{Chapman1952}
\begin{equation}
    \label{eq:eta_0_chapman}
    \eta^0_i = \frac{5}{16\sigma_\mathrm{cross}^2 \Omega_i^{*(2, 2)}} \left ( \frac{m_i}{\beta\pi} \right )^{1/2}~,
\end{equation}
where the cross sectional diameter $\sigma_\mathrm{cross}$ and the collision cross section integral $\Omega_i^{*(2, 2)}$ depend on the interaction potential.
For hard-sphere fluids, $\sigma^2_\mathrm{cross}\Omega_i^{*(2, 2)}=\sigma_i^2$.

When the fluid interactions are not rigid, $\sigma_\mathrm{cross}$ is typically chosen to equal some characteristic distance of the interaction potential, and  
$\Omega_i^{*(2, 2)}$ can then, at least in principle, be calculated by numerically integrating the expression given for $\Omega_i^{*(2, 2)}$ by Chapman in Ref.~\cite{Chapman1952}.
However, 
such an approach is not guaranteed to provide a satisfactory level of accuracy when predicting the low-density viscosity of fluids with complex interactions. 
For example, it cannot take into account the impact of any additional degrees of freedom such as rotation and internal vibration.
In practice, in such cases, $\Omega_i^{*(2, 2)}$ can alternatively be treated as a fit parameter, or determined based on numerical data for the low-density viscosity of a fluid. The resulting effective value $\Omega_i^\mathrm{eff}$ obtained by such a procedure will typically differ from the exact integral expression, due to the limitations of the assumptions made in Chapman's theory.

The remainder of this section introduces some important notation and provides a summary of the Enskog expression for the shear viscosity of dense hard-sphere fluids, along with some extensions of Enskog's work that are relevant for the development of the approach presented in Sec.~\ref{sec:Dipolar_DHS_theory}. This includes DHS theory, which extends the Enskog expression to pure DHS fluids~\cite{Pousaneh2}, as well as the Enskog-Thorne expression for the shear viscosity of binary hard-sphere mixtures.

\subsection{Chapman-Enskog theory}

Chapman-Enskog theory gives the expression for the density dependent shear viscosity of pure hard sphere fluids of type $i$ as~\cite{Chapman1952} 
\begin{equation}
    \label{eq:Enskog}
    \eta = \eta^0_i \left ( \frac{1}{\chi_i} + \alpha_i \rho + \frac{1}{4}\alpha_i^2\rho^2\chi_i \right ) + \frac{3}{5}\bar{\omega}_i~,
\end{equation}
where
\begin{equation}
\label{eq:alpha_HS}
    \alpha_i = \frac{8}{15}\pi\sigma_i^3~,
\end{equation}
and 
\begin{equation}
    \label{eq:omega_bar}
    \bar\omega_i = \frac{5}{\pi}\eta^0_i\alpha_i^2\rho^2\chi_i~.
\end{equation}
The quantity $\chi_i$ is the pair distribution function (PDF) $g(r)$ evaluated at the point of hard-sphere contact $\chi_i = \lim_{r \to \sigma_i^+} g(r)$, and contains the non-linear density dependence of the collision rate at intermediate and high densities, which is proportional to $\rho\chi_i$.

\subsection{Dipolar hard sphere theory}
\label{sec:pDHS}

To apply Chapman-Enskog theory to pure DHS fluids, the main assumption needed by Pousaneh and de Wijn~\cite{Pousaneh2}
was that the majority of momentum transfer between interacting particles will be the result of collisions between the hard cores.
The dipolar interactions only modify the hard-core collision rate.
Thus, $\chi_i = \chi_i^\mathrm{DHS}$, where $\chi_i^\mathrm{DHS}$ is the PDF contact value for the DHS fluid. Additionally, $\alpha_i$ and $\bar{\omega}_i$ are calculated based on the hard-core diameter $\sigma_i$. 
In the rest of this paper, the resulting theory for the shear viscosity of the pure DHS fluid, including the expression derived by Pousaneh and de Wijn for $\chi_i^\mathrm{DHS}$ discussed below, will be referred to as pure DHS (pDHS) theory.

To derive an expression for
$\chi_i^\mathrm{DHS}$, Pousaneh and de Wijn~\cite{Pousaneh2} made use of a relation for DHS fluids which relates $\chi_i^\mathrm{DHS}$ to the compressibility factor $Z = \beta P/\rho$ and the potential energy associated with the dipolar interactions $u_\mathrm{pot}$, both of which can be calculated based on a sufficiently accurate expression for the free energy of the pure DHS fluid. 
For this, pDHS theory relies on logarithmic free energy (LFE) theory derived by Elfimova~\textit{et al.}~\cite{Elfimova2012, Vtulkina2016}, which re-sums the dipolar contribution to the DHS virial series as a logarithmic function, motivated by the functional form $\beta F/N = - \ln{Q^{1/N}}$.

Pure DHS theory provides a good description of the density-dependent behavior of the shear viscosity for simulated DHS fluids using an effective dipole coupling constant~$\lambda_i^\mathrm{eff}$, which replaces the true value $\lambda_i$ that appears in the expression for $\chi^\mathrm{DHS}_i$. The values for $\lambda_i^\mathrm{eff}$ reported by Pousaneh and de Wijn are considerably lower than the true dipolar coupling strength, which is unsurprising, given that DHS fluids violate several of the assumptions made by Boltzmann and Enskog. 
It is assumed that these effective values account for the way in which rotational degrees of freedom and other things not explicitely included, and how they modify the momentum exchange rate between colliding DHS particles.

For the low density limit, pDHS theory sets $\sigma_\mathrm{cross} = \sigma_i$. The effective collision cross section integral is treated as an additional fit parameter $\Omega_{i}^\mathrm{pDHS}$. It is worth noting that, because DHS particles form chain like clusters at low density~\cite{Weis2006, Wei2011, Pousaneh2, Fjeldstad2023}, $\chi^\mathrm{DHS}_i$ does not go to unity in the low density limit. As a consequence, $\Omega_{i}^\mathrm{pDHS} \neq \Omega_{i}^\mathrm{eff}$. Rather, it follows from comparison of pDHS theory with the standard Enskog expression for hard spheres that 
$\Omega_{i}^\mathrm{pDHS} = \Omega_{i}^\mathrm{eff} / \lim_{\rho \to 0} (\chi^\mathrm{DHS}_i|_{\lambda_i = \lambda_i^\mathrm{eff}})$.

\subsection{Enskog-Thorne theory for mixtures}
\label{sec:E-T-theory}

The shear viscosity expression given by Eq.~(\ref{eq:Enskog}) was extended by Thorne~\cite{Chapman1952} for multiple components. The original expression for the shear viscosity of binary mixtures can be found in the book by Chapman and Cowling~\cite{Chapman1952} and is further expanded upon in Di Pippo~\textit{et al.}~\cite{di_pippo} where it is combined with modified Enskog theory~\cite{Hanley1972} and applied to real gases. Enskog-Thorne theory is known to be inconsistent with the Onsager reciprocal relations~\cite{Barajas1973}. However, this issue does not affect the expression for the shear viscosity~\cite{di_pippo} and so it remains reliable for our purpose.

Although available from other sources, it is useful to include the full expression for the shear viscosity of binary hard sphere mixtures here.
Before presenting Thorne's expression, we first go through the mixing parameters that it requires.
For each parameter or quantity needed to obtain the viscosity of the pure fluids, we now also need mixing parameters related to the interaction between any two components in the mixture.

The low-density mixing viscosity $\eta^0_{ij}$ is the viscosity of a hypothetical pure gas consisting of particles with mass $m_{ij} = 2m_im_j/(m_i + m_j)$ and $\sigma_\mathrm{cross} = \sigma_{ij}$, given by
\begin{equation}
    \label{eq:eta_0_mixtures}
    \eta^0_{ij} = \frac{5}{16\sigma_{ij}^2 \Omega^{*(2, 2)}_{ij}} \left ( \frac{m_{ij}}{\beta\pi} \right )^{1/2}~,
\end{equation}
where $\Omega^{*(2, 2)}_{ij}$ is the collision cross section integral of a pure fluid for which all inter-particle interactions are equal to the interaction between the relevant $i,j$-species.
As for pure fluids, $\Omega^{*(2, 2)}_{ij}$ could, if necessary, be treated as a fit parameter $\Omega^\mathrm{eff}_{ij}$.

The contact values $\chi_{ij}$ are defined in terms of the partial PDFs $g_{ij}(r)$, so that $\chi_{ij} = \lim_{r \to \sigma_{ij}^+} g_{ij}(r)$. By symmetry, $\chi_{ij} = \chi_{ji}$. In the mixture, $\chi_{ij}$ depends on the density as well as the composition, even for like-like interactions.

Finally, we need the mixture coefficients $\alpha_{ij}$ and $\bar{\omega}_{ij}$.
The former can be obtained by substituting $\sigma_{ij}$ for $\sigma_i$ in Eq.~(\ref{eq:alpha_HS}), giving $\alpha_{ij} = 8\pi\sigma_{ij}^3/15$.
The latter is calculated by substituting the $ij$-mixing parameters into Eq.~(\ref{eq:omega_bar}),
giving $\bar{\omega}_{ij} = 5\eta_{ij}^0\alpha_{ij}^2\rho^2\chi_{ij}/\pi$.

The shear viscosity of binary hard-sphere mixtures with particles species denoted type 1 and type 2 is given by Enskog-Thorne theory as
\begin{align}
\label{eq:Thorne}
    \eta_\mathrm{mix} =\frac{ {y_1^2H_{22}} + {y_2^2H_{11}} - {2y_1y_2H_{12}}{}}{ H_{11}H_{22} - {H_{12}^2} } + \frac{3}{5}\bar{\omega}_\mathrm{mix}~,
\end{align}
where $y_1$, $y_2$, $H_{11}$, $H_{22}$, $H_{12}$ and $\bar{\omega}_\mathrm{mix}$ can all be expressed in terms of the mixing parameters defined above. For $y_1$ and $y_2$ we have,
\begin{equation}
\label{eq:y1}
    y_1 = x_1 \left( 1 + \frac{1}{2}x_1\alpha_{11}\chi_{11}\rho + \frac{m_2}{m_1 + m_2}x_2\alpha_{12}\chi_{12}\rho \right)~,
\end{equation}
and
\begin{equation}
\label{eq:y2}
    y_2 = x_2 \left( 1 + \frac{1}{2}x_2\alpha_{22}\chi_{22}\rho + \frac{m_1}{m_1 + m_2}x_1\alpha_{12}\chi_{12}\rho \right)~.
\end{equation}
For $H_{11}$, $H_{22}$, and $H_{12}$ we have 
\begin{equation}
    \label{eq:H11}
    H_{11} = \frac{x_1^2\chi_{11}}{\eta^0_{11}} + \frac{2x_1x_2\chi_{12}}{\eta^0_{12}}\frac{m_1m_2}{(m_1 + m_2)^2} \left( \frac{5}{3A^*_{12}} + \frac{m_2}{m_1} \right)~,
\end{equation}
\begin{equation}
    \label{eq:H22}
    H_{22} = \frac{x_2^2\chi_{22}}{\eta^0_{22}} + \frac{2x_1x_2\chi_{12}}{\eta^0_{12}}\frac{m_1m_2}{(m_1 + m_2)^2} \left( \frac{5}{3A^*_{12}} + \frac{m_1}{m_2} \right)~,
\end{equation}
and 
\begin{equation}
    \label{eq:H12}
    H_{12} = -\frac{2x_1x_2\chi_{12}}{\eta^0_{12}}\frac{m_1m_2}{(m_1 + m_2)^2} \left( \frac{5}{3A^*_{12}} - 1 \right)~,
\end{equation}
where the constant $A^*_{12}$ is the dimensionless ratio of the diffusive and viscous collision integrals. When only hard sphere interactions are present, $A^*_{12}$ is equal to unity. Finally, for $\bar{\omega}_\mathrm{mix}$ we have
\begin{equation}
    \label{eq:omega_mix}
    \bar{\omega}_\mathrm{mix} = x_1^2\bar{\omega}_{11} + 2x_1x_2\bar{\omega}_{12} + x_2^2\bar{\omega}_{22}~.
\end{equation}

\section{Shear viscosity of dense binary dipolar hard sphere fluid mixtures}
\label{sec:Dipolar_DHS_theory}
We now turn to binary mixtures of DHS, and describe our approach for predicting the composition dependent shear viscosity of binary DHS fluid mixtures, which we will refer to as binary mixture DHS (bmDHS) theory.
We make similar assumptions as in pDHS theory.
We assume that the majority of momentum exchange between particles is the result of collisions between the hard-sphere cores and so $\chi_{ij} = \chi^\mathrm{DHS}_{ij}$, where $\chi^\mathrm{DHS}_{ij}$ denotes the partial PDF contact values for the dipolar mixture. Similarly, we calculate $\alpha_{ij}$ using $\sigma_{ij}$, the same as for mixtures of true hard spheres.

In order to apply Enskog-Thorne theory
we require suitable expressions for $\chi^\mathrm{DHS}_{ij}$.
These must take into account the impact of composition on the fluid structure associated with the dipolar interaction.
In the present work, we make use of the results of Novak \textit{et al.}~\cite{Novak2013}, who provide a first order term in the density expansion for the partial PDFs of binary DHS fluids.  We then make two physically motivated corrections to improve their performance at intermediate and high density.
This is done in Sec.~\ref{sec:PDF_DHS}.

As a consequence of sharing the same set of assumptions as pDHS theory, bmDHS theory will need to rely on effective dipolar coupling constants $\lambda^\mathrm{eff}_{ij}$, which differ significantly from the true values, for the density dependent behavior. 
The theory will also rely on effective cross section integrals $\Omega^\mathrm{bmDHS}_{ij}$.
Similarly to pDHS theory, $\chi^\mathrm{DHS}_{ij}$ does not go to unity in the low density limit and so 
\begin{equation}
\label{eq:Omega_bmDHS}
    \Omega^\mathrm{bmDHS}_{ij} = \frac{\Omega^\mathrm{eff}_{ij}}{\lim_{\rho \to 0} (\chi^\mathrm{DHS}_{ij}|_{\lambda_{ii} = \lambda^\mathrm{eff}_{ii}, \lambda_{jj} = \lambda^\mathrm{eff}_{jj}})}~.
\end{equation}
The values of both $\lambda^\mathrm{eff}_{ij}$ and $\Omega^\mathrm{bmDHS}_{ij}$ depend only on the interactions and properties of the relevant species. Crucially, they do not depend on the composition of the mixture.
Thus, when $i = j$, the values for these parameters can be obtained by fitting bmDHS theory in the pure limit to shear viscosity data for pure DHS fluids.
These like-like interaction parameters can then be used to extract the interaction parameters associated with the $1,2$-interaction, $\lambda^\mathrm{eff}_{12}$ and $\Omega^\mathrm{bmDHS}_{12}$, which are discussed in Sec.~\ref{sec:1-2-params}.

\subsection{Partial pair distribution functions for binary DHS fluids}
\label{sec:PDF_DHS}

In this work, we take an approach that uses expressions for $\chi^\mathrm{DHS}_{ij}$ obtained from cluster theory~\cite{Novak2013}, and further improve them using a re-summation approach.
Cluster theory gives an expression for partial PDFs in terms of a power series in density.
For a binary DHS fluid mixture this gives
\begin{equation}
\label{eq:g_r_power_series}
    g_{ij}(\vec{r}) = \sum_{n = 2}^\infty B^\mathrm{DHS}_{n,ij}(\vec{r}) \rho^{n - 2}~,
\end{equation}
where $B^\mathrm{DHS}_{n,ij}(\vec{r})$ are cluster expansion coefficients. In the present work, it is assumed that no external field is present, meaning the binary fluid mixture is isotropic and we can safely ignore the impact of orientation on the partial PDF. 
Furthermore, as we are only interested in the hard-core contact value for which $r = \sigma_{ij}$, Eq.~(\ref{eq:g_r_power_series}) simplifies to
\begin{equation}
\label{eq:chi_power_series}
    \chi_{ij}^\mathrm{DHS} = \sum_{n = 2}^\infty B^\mathrm{DHS}_{n,ij} \rho^{n - 2}~,
\end{equation}
where $B^\mathrm{DHS}_{n,ij} = B^\mathrm{DHS}_{n,ij}(\vec{r})|_{|\vec{r}|=\sigma_{ij}}$. 

Expressions for the first few coefficients already exist in the literature for DHS fluids, both for pure fluids~\cite{Elfimova2010}, and for binary fluid mixtures~\cite{Novak2013, Nekhoroshkova2014}. 
In addition to expanding $g_{ij}(r)$ in $\rho$, such expressions treat 
the individual coefficients $B^\mathrm{DHS}_{n,ij}(r)$ as power series expansions in $\lambda_{ij}$.
Novak~\textit{et al.}~\cite{Novak2013} derived expressions for $B^\mathrm{DHS}_{ij,2}(r)$ and $B^\mathrm{DHS}_{ij,3}(r)$ (i.e. up to linear terms in the density) for the binary dipolar hard sphere fluid mixture that consider terms up to the fourth power in $\lambda_{ij}$.
To our knowledge, no additional terms beyond the work of Novak~\textit{et al.}\ have been derived for the partial PDF cluster expansion for binary DHS fluid mixtures. 
Calculation of additional higher order terms in either $\rho$ or $\lambda_{ij}$ is not practical for binary DHS fluid mixtures due to the enormous configurational complexity associated with these higher order terms.

Due to the long-range nature of dipolar interactions, the cluster series in Eqs.~(\ref{eq:g_r_power_series}) and~(\ref{eq:chi_power_series}) converge very slowly in the density and so exhibit limited accuracy at elevated density when based on the Novak~\textit{et al.} expressions. Furthermore, truncation of $B^\mathrm{DHS}_{n,ij}(r)$ beyond the fourth power in $\lambda_{ij}$ limits the accuracy of Novak~\textit{et al.}\ for strongly interacting dipoles. 

\begin{figure}
    \centering
    \includegraphics[width=0.95\columnwidth]{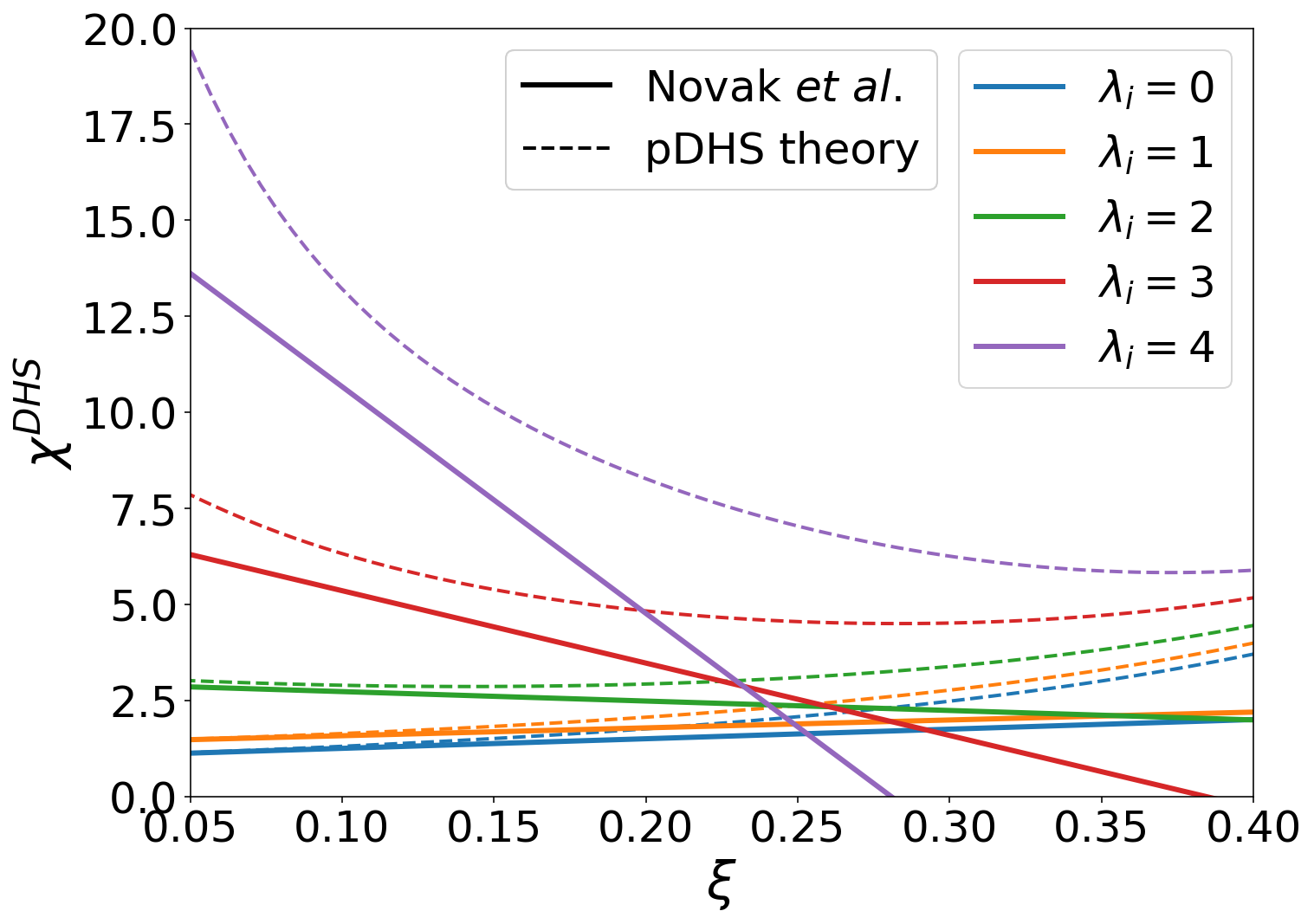}
    \caption{Prediction for $\chi^\mathrm{DHS}_i$ using the cluster expansion derived by Novak~\textit{et al.}~\cite{Novak2013} in the pure limit (full lines) along with, for comparison, the equivalent prediction using pDHS theory for pure fluids~\cite{Pousaneh2} (dashed lines). 
    At low density, as long as $\lambda_i \lesssim 2$, the Novak~\textit{et al.} prediction is very similar to pDHS theory. However, the higher-order terms in the density become significant, and Novak~\textit{et al.}'s expansion for mixtures fails. 
    }
    \label{fig:Comparison_lambda_Novak}
\end{figure}

The poor convergence of the virial expansion can be seen clearly in Fig.~\ref{fig:Comparison_lambda_Novak}, where we compare the pure limit of Novak \textit{et al.}'s prediction, 
$\chi^\mathrm{Novak}_{ii}|_{x_j = 0}$,
with the equivalent pDHS prediction which relies on LFE theory. 
When both $\xi \lesssim 0.15$ and $\lambda_i \lesssim 2$, Novak~\textit{et al.}\ is largely in agreement with LFE theory. However, as the volume fraction increases, the two predictions rapidly diverge, even when $\lambda_{ii}$ is small. Especially for strong dipoles, i.e. $\lambda_{ii} \gtrsim 3$, the higher order terms in the density become significant, and Novak~\textit{et al.}'s expansion for mixtures fails, demonstrating that the convergence in the density expansion is insufficient.

Although the expressions based on a straight-forward cluster expansion are not sufficiently accurate for our purpose, they do provide a useful starting point for the development of more accurate expressions for $\chi_{ij}^\mathrm{DHS}$.
In this work, we start from Novak~\textit{et al.}'s results and improve convergence in the density expansion by making two physics-motivated corrections.
The first is a correction to the hard-sphere contribution based on extensive knowledge in the literature about the correlations in mixtures of hard spheres without dipoles (for a review on hard-sphere theories, see Refs.~\cite{Santos2020, Mulero2008}).
The second correction is based on a re-summing of the residual dipolar contribution, inspired by LFE theory, using a well-chosen, physically motivated, function as detailed below.

\subsubsection{Correction based on hard-sphere theory}
For the first correction, we split $\chi^\mathrm{DHS}_{ij}$ into two terms by separating out the hard-sphere contribution $\chi^\mathrm{HS}_{ij} = \chi^\mathrm{DHS}_{ij}|_{\lambda_{ii} = \lambda_{jj} = 0}$ giving
\begin{equation}
\label{eq:chi_as_a_sum}
    \chi^\mathrm{DHS}_{ij} = \chi^\mathrm{HS}_{ij} + \Delta \chi^\mathrm{DHS}_{ij}~,
\end{equation}
where the residual dipolar contribution $\Delta \chi^\mathrm{DHS}_{ij} = \chi^\mathrm{DHS}_{ij} - \chi^\mathrm{HS}_{ij}$ measures the degree to which $\chi^\mathrm{DHS}_{ij}$ deviates from the hard sphere fluid mixture.
We can then correct the HS contribution in Eq.~(\ref{eq:chi_as_a_sum}) by replacing it with a more accurate expression for $\chi_{ij}^\mathrm{HS}$ such as the one from
Ref.~\cite{BMCSL}, denoted BMCSL (Boubl\'ik, Mansoori, Carnahan, Starling and Leland) theory, so that
\begin{equation}
\label{eq:chi_as_a_sum_BMCSL}
    \chi^\mathrm{DHS}_{ij} = \chi^\mathrm{BMCSL}_{ij} + \Delta \chi^\mathrm{DHS}_{ij},
\end{equation}
where
\begin{equation}
    \label{eq:CS_mixtures}
    \begin{split}
    \chi_{ij}^\mathrm{BMCSL} & = \frac{1}{1 - \xi_{(3)}}
                              + \frac{3\xi_{(2)}}{(1 - \xi_{(3)})^2}\frac{\sigma_{i} \sigma_{j}}{\sigma_i + \sigma_j} \\
                             &\qquad\qquad\qquad + \frac{2\xi_{(2)}^2}{(1 - \xi_{(3)})^3} \left ( \frac{\sigma_{i} \sigma_{j}}{\sigma_i + \sigma_j} \right )^2~,
    \end{split}
\end{equation}
with $\xi_{(\ell)} = \pi \rho \sum_i x_i\sigma_i^\ell/6$.

\subsubsection{Re-summing the Dipolar Contribution}
\label{sec:re-summing}

We must also further refine the expression for $\Delta \chi^\mathrm{DHS}_{ij}$ in order to improve the convergence of the density expansion and avoid unphysical negative collision rates.
As mentioned above, obtaining more coefficients in the expansion is extremely time consuming.
Instead, we introduce two physically-motivated conditions, and then re-sum the expansion in accordance.

The physical basis for our resumming approach is the physical observation that all correlation functions and probabilities in equilibrium are governed by the Boltzman factor, which is an exponential function of the energy.
Any correlation function is an ensemble average, and takes the shape of an average of exponential functions of energies.

We first make the observation that this means that the ensemble average of the energy of the dipolar interaction is always net negative. 
Therefore, in equilibrium, polar particles are on average found closer together than if they were not polar, and $\chi_{ij}^\mathrm{DHS} > \chi_{ij}^\mathrm{HS}$.
We must therefore ensure that not only $\chi_\mathrm{ij}>0$, but also $\Delta\chi_{ij}^\mathrm{DHS} > 0$ for all $\rho$.
With the truncated virial expansion of Novak~\textit{et al.}\ this condition of positivity is clearly not generally satisfied (see Fig.~\ref{fig:Comparison_lambda_Novak}).

Motivated by the general form of probabilities in equilibrium, we expand the residual cluster series for $\Delta\chi_{ij}^\mathrm{DHS}$
as an exponential function,
\begin{equation}
\label{eq:chi_D_exp}
    \Delta \chi^\mathrm{DHS}_{ij} = \exp{\left [ \sum_{n = 0}^\infty A_{n,ij} \rho^n \right ]}~,
\end{equation}
where $A_{n,ij}$ are the new expansion coefficients.
We thus replace a complicated average of exponential functions over all coordinates but two, by an exponential function of only two coordinates, effectively neglecting any effects of 3-body correlations, but keeping contributions to the higher order terms from 2-body correlations.
This also guarantees that $\Delta\chi_{ij}^\mathrm{DHS} > 0$ will always be satisfied. Thus, we write
While the exponential function is to some degree an arbitrary choice, it is physically motivated, and similar to what is done in LFE theory with a logarithmic form when re-summing the residual component of the virial free energy series~\cite{Elfimova2012}.

The coefficients $A_{n,ij}$ can be extracted by matching the coefficients in the lowest orders in Novak's density expansion with the density expansion of the exponential function.
We define the residual coefficients in Novak's expansion using
\begin{equation}
\label{eq:chi_D_power}
    \Delta \chi^\mathrm{DHS}_{ij}  = \sum_{n = 2}^\infty \Delta B^\mathrm{DHS}_{n,ij} \rho^{n - 2}~,
\end{equation}
where $\Delta B^\mathrm{DHS}_{n,ij} = B^\mathrm{DHS}_{n,ij} - B^\mathrm{DHS}_{n,ij}|_{\lambda_{11} = \lambda_{22} = 0}$.
Only two coefficients in this expansion are known from Novak's cluster expansion~\cite{Novak2013}, for the constant and linear terms.
Expanding Eq.~(\ref{eq:chi_D_exp}) up to the same order gives
\begin{equation}
\label{eq:chi_D_exp_taylor}
    \Delta \chi^\mathrm{DHS}_{ij} = \exp{[A_{0,ij}]} + \exp{[A_{0,ij}]}A_{1,ij}\rho + \mathcal{O}(\rho^2)~.
\end{equation}
Comparing and matching terms in  Eq.~(\ref{eq:chi_D_exp_taylor}) with terms in Eq.~(\ref{eq:chi_D_power}) we find that
\begin{equation}
    A_{0,ij} = \ln{\Delta B^\mathrm{DHS}_{ij,2}}~, \quad A_{1,ij} = \frac{\Delta B^\mathrm{DHS}_{ij,3}}{\Delta B^\mathrm{DHS}_{ij,2}}~.
\end{equation}
Inserting these coefficients into  Eq.~(\ref{eq:chi_D_exp}), truncating beyond $n=1$, and combining with Eq.~(\ref{eq:chi_as_a_sum_BMCSL}), we end up with the final,
\begin{equation}
\label{eq:chi_as_a_sum_BMCSL_resum}
    \chi^\mathrm{DHS}_{ij} = \chi^\mathrm{BMCSL} + \exp{\left [ \ln{\Delta B^\mathrm{DHS}_{2,ij}} + \frac{\Delta B^\mathrm{DHS}_{3,ij}}{\Delta B^\mathrm{DHS}_{2,ij}}\rho + \mathcal{O}(\rho^2) \right ]}~,
\end{equation}
which is the expression that we use for $\chi^\mathrm{DHS}_{ij}$ in this work.
If at any point in the future more coefficients become available from the cluster expansion, this expression could be refined to include those higher orders.

\subsubsection{Comparison}
\begin{figure}
    \centering
    \includegraphics[width=0.95\columnwidth]{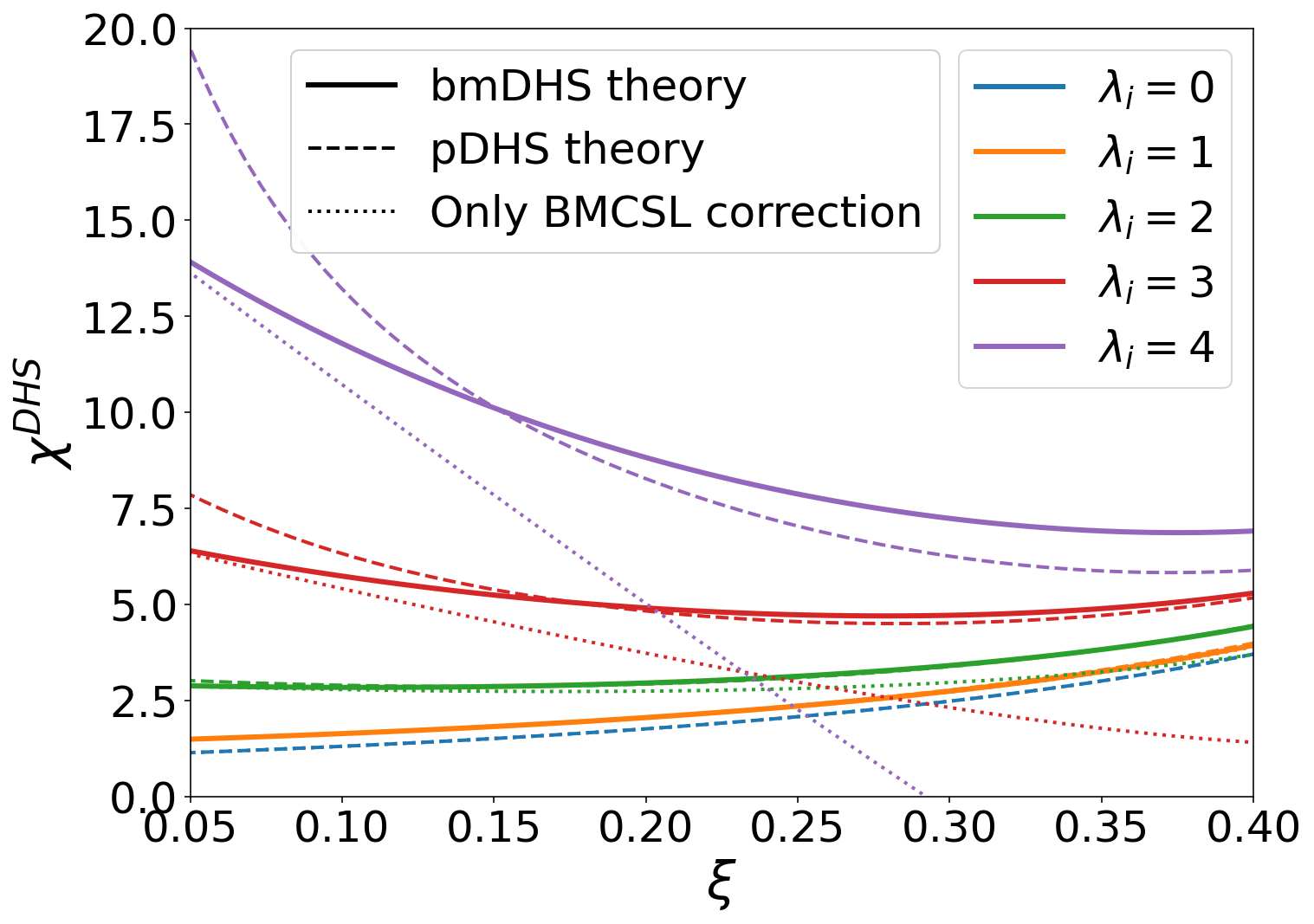}
    \caption{Prediction for $\chi^\mathrm{DHS}_i$ in the pure limit: fully corrected expression for $\chi_{ij}$, denoted bmDHS theory (full lines), the equivalent prediction using pDHS theory for pure fluids~\cite{Pousaneh2} (dashed lines), and only the BMCSL correct for the hard-sphere contribution (dotted lines). Although some improvement can be seen in the latter, especially for weak dipoles, it still fails when $\lambda_i = 4$. In contrast, bmDHS theory, which uses both corrections, shows far better agreement with pDHS theory, although very strong dipoles still present an quantitative issue. Crucially, bmDHS theory satisfies the sanity requirement that $\chi^\mathrm{DHS}_i > 0$ for all densities.}
    \label{fig:Comparison_lambda_resum}
\end{figure}

Figure~\ref{fig:Comparison_lambda_resum} compares the prediction for $\chi^\mathrm{DHS}_{ii}|_{x_i = 0}$ calculated using  Eq.~(\ref{eq:chi_as_a_sum_BMCSL_resum}) (full lines), with the same equivalent pDHS prediction used for comparison in Fig.~\ref{fig:Comparison_lambda_Novak} (dashed lines). 
The figure also includes the prediction for $\chi^\mathrm{DHS}_{ii}|_{x_i = 0}$ made using Eq.~(\ref{eq:chi_as_a_sum_BMCSL}) (dotted lines) which includes the hard-sphere BMCSL correction, but not the re-summation of the dipolar residual.
We do not include any numerical results, because our numerical approach does not allow for calculating an accurate contact value of the partial PDF.  We rely on the pseudo hard-sphere potential for doing molecular dynamics simulations as detailed in Sec.~\ref{sec:SimSetUp}, which smears out the initial peak of the partial PDFs compared to a true hard-sphere fluid.

Because the higher order hard-sphere contributions are taken into account in a better way, Eq.~(\ref{eq:chi_as_a_sum_BMCSL}) performs better than just the simple cluster expansion up to linear order.  However, it is still severely limited by the missing higher order terms in the dipolar contribution, and still predicts $\chi^\mathrm{DHS}_i < 0$ when $\lambda_{i}$ is sufficiently large.

The prediction using Eq.~(\ref{eq:chi_as_a_sum_BMCSL_resum}), including both the accurate hard-sphere BMCSL correction and the re-summation to improve the dipolar contribution, are generally well-behaved.
For most dipole moments, they are qualitatively and quantitatively similar to pDHS theory. 
Nevertheless, for very strong dipoles there are some quantitative differences, especially at low densities, as could be expected from the fact that the higher-order terms in the expansion in the coupling constants are truncated earlier than in LFE theory.

\subsection{Dipolar mixing parameters}
\label{sec:1-2-params}

We estimate the dipolar coupling constant $\lambda_{12}^\mathrm{eff}$, by
inserting $\mu_{1} = \sqrt{4\pi\epsilon_0\sigma_1^3\lambda_{11}/\beta}$ and $\mu_{2} = \sqrt{4\pi\epsilon_0\sigma^3_2\lambda_{22}/\beta}$ into Eq.~(\ref{eq:lambda_ij}), and substituting in the effective dipolar coupling constants. This gives
\begin{align}
    \label{eq:lambda_eff_ij}
    \lambda^\mathrm{eff}_{12}  = \left ( \frac{2\sqrt{\sigma_1\sigma_2}}{\sigma_1 + \sigma_2} \right )^3\sqrt{\lambda^\mathrm{eff}_{11}\lambda^\mathrm{eff}_{22}}~.
\end{align}

In principle, $\Omega^\mathrm{bmDHS}_{12}$ can simply be treated as a mixture-dependent fit parameter, similar to its pure-fluid counterparts. However, in the present work, we propose a route for estimating $\Omega^\mathrm{bmDHS}_{12}$ based on only pure fluid data.
According to Enskog-Thorne theory, the effective parameter $\Omega^\mathrm{eff}_{12}$ should take the value that $\Omega^\mathrm{eff}_{i}$ would take for a hypothetical fluid for which all interactions are equal to the interaction between the $1,2$-species. Because bmDHS theory relies on effective dipolar coupling constants to quantify the dipolar interaction, it follows from Eq.~(\ref{eq:Omega_bmDHS}) that 
\begin{align}
    \label{eq:omega_eff_12}
    \Omega^\mathrm{bmDHS}_{12}  = \frac{\Omega(\lambda^\mathrm{eff}_{12})
    }{\lim_{\rho \to 0} (\chi^\mathrm{DHS}_{ij}|_{\lambda_{ii} = \lambda^\mathrm{eff}_{11}, \lambda_{jj} = \lambda^\mathrm{eff}_{22}})}~,
\end{align}
where, for convenience, we introduce the function $\Omega(\lambda) = \Omega^\mathrm{eff}_{ij}|_{\lambda^\mathrm{eff}_{ij} = \lambda}$.
Applying Eq.~(\ref{eq:omega_eff_12}) requires knowledge of $\Omega(\lambda)$.
The data presented for pure fluids later in Sec.~\ref{sec:pure_DHS} strongly suggests that
an expansion in $\lambda^\mathrm{eff}_{ij}$ is a suitable approach for the range of parameters investigated in this work.
The reader is referred to that section for a more thorough discussion.

Lastly, we have $A^*_{12}$.
When applying Enskog-Thorne theory to binary mixtures of real non-polar and mostly noble gases, Di Pippo~\textit{et al.}~\cite{di_pippo} found that instead of $A^*_{12}$ equal to unity, $A^*_{12} = 1.1$ provided reasonable results in most cases. However, the exact value of $A^*_{12}$ was found by Di Pippo~\textit{et al.} to have only a limited impact on the prediction. We therefore simply use the same value of 1.1 here.

\section{Simulation Setup}
\label{sec:SimSetUp}

We perform numerical experiments on the system described in Sec.~\ref{sec:PhysMod} using Molecular Dynamics (MD) simulations in LAMMPS~\cite{LAMMPS}. 
The general setup, the inter-particle interactions and the specific simulation parameters that we use are all similar to that of Ref.~\cite{Pousaneh2}, and the technical setup and viscosity calculations in LAMMPS are similar to that described in Ref.~\cite{Fjeldstad2023}.

The simulation parameters, predictions and numerical results are all expressed in reduced LJ units which are denoted by the presence of a superscript asterisk. Physical quantities are given in units relative to a reference mass $M$, diameter $\sigma$ and energy scale $\epsilon$, as well as the Boltzmann constant, $k_\mathrm{B}$. 
The most relevant reduced units in the present work are mass $m^* = m/M$, distance $r^* = r/\sigma$, time $\tau^* = \tau(\epsilon / m \sigma^2)^{1/2}$, 
temperature $T^* = k_\mathrm{B}T/\epsilon$,
dipole moment $\mu^* = \mu / (4 \pi \epsilon_0 \sigma^3 \epsilon)^{1/2}$, and viscosity $\eta^* = \eta \sigma^3 / (\epsilon \tau)$.

\subsection{Particle properties}
\label{sec:sim_params}

DHS particles are categorized based on the size of their hard-sphere core. We include two distinct categories for investigation: 1) small particles for which the hard-sphere diameter $\sigma^*_i = 1$ and the mass $m^*_i = 1$, and 2) large particles for which the hard-sphere diameter $\sigma^*_i = 1.5$ and the mass $m^*_i = 3.375$. 
The large particle mass is chosen to scale with the volume of the core. 
In LAMMPS, the moment of inertia is controlled by choosing a mass distribution diameter separate from the hard sphere interaction. The particle mass is then uniformly distributed across the volume of the associated sphere. In the present work, we set this diameter equal to $0.316\sigma^*_i$. The resulting moment of inertia $I^*_i =  0.0099856 m^*_i {\sigma^*_i}^2$ is similar to that investigated by Ref.~\cite{Pousaneh2} when $\sigma_i^* = 1$ and $m^*_i = 1$. In our setup, the moment of inertia scales with $m^*_i{\sigma^*_i}^2$, so that the large particles have a moment of inertia that is $3.375 \cdot 1.5^2 \approx 7.6$ times greater than that of the small particles.

We consider two classes of mixtures. First are mixtures that are symmetric with respect to the size and mass of the component hard-cores, but asymmetric with respect to their dipolar coupling strengths.
Specifically, we set $\sigma^*_1 = \sigma^*_2 = 1$ and $m^*_1 = m^*_2 = 1$. 
We choose the second component to have the largest dipole moment.
Secondly, we consider mixtures that are fully asymmetric and use the convention that component 1 is the smaller component.
In both types of mixtures, we consider values for $\lambda_{ii}$ in the range $1 \leq \lambda_{ii} \leq 4$. This is the same range investigated for pure fluids by Ref.~\cite{Pousaneh2}.
In the fully asymmetric mixtures, we explore both combinations, i.e. the larger particle can have either the larger dipole moment or the smaller dipole moment.

All mixtures are simulated across the full range of possible compositions $0 \leq x_2 \leq 1$.
Note that $x_2 = 0$ and $x_2 = 1$ correspond to the pure fluids which are included for the purpose of extracting effective parameters $\lambda^\mathrm{eff}_{ii}$ and $\Omega^\mathrm{bmDHS}_{ii}$ as well as verifying the limiting behavior of our theory as described in Sec.~\ref{sec:pure_DHS}.

\subsection{Inter-particle interactions}
Because LAMMPS is designed for simulating smooth interactions, the hard-sphere component of the interaction is approximated using the pseudo hard sphere (PHS) potential developed by Jover~\textit{et al.}~\cite{Jover2012} given by the cut and shifted Mie-potential,
\begin{equation}
    \label{eq:PHS}
    u^\mathrm{PHS}_{ij} = \begin{cases} 50 \left( \frac{50}{49} \right)^{49} \epsilon \left[ \left( \frac{\sigma_{ij}}{r} \right)^{50} - \left( \frac{\sigma_{ij}}{r} \right)^{49} \right] + \epsilon, & r > \frac{50}{49}\sigma_{ij} \\
    0, & r \le \frac{50}{49}\sigma_{ij}
    \end{cases}~,
\end{equation}
at a reduced temperature $T^* = 1.5$.
The PHS potential has successfully been applied to calculations of viscosity for both pure hard-sphere fluids~\cite{Pousaneh1} and DHS fluids~\cite{Pousaneh2} from MD simulations.
We have tested the PHS potential for asymmetric binary mixtures of hard spheres with no polar interaction, and compared the viscosity of these mixtures to Enskog-Thorne theory for hard spheres and found no issues as long as $\sigma_i \leq 2$ for both species.

For computational efficiency, the long-range dipolar interactions are calculated using Ewald summation as implemented in the KSPACE package of LAMMPS for a particle-particle particle mesh dipole solver (pppm/dipole)~\cite{Cerda2008}. The crossover value $r^*_\mathrm{cut}$ determines the range at which direct calculations for the dipolar interactions are performed. 
We use $r^*_\mathrm{cut} = 5.0$ when $\xi \leq 0.15$, $r^*_\mathrm{cut} = 4.5$ when $0.2 \leq \xi \leq 0.25$, and $r^*_\mathrm{cut} = 4.0$ when $\xi \geq 0.30$.
The Ewald accuracy parameter used to determine mesh size in conjunction with the tabulated cutoffs is $0.05$. 
We confirmed that this accuracy is sufficient by comparing against more accurate systems for a small sample of test parameters. The resulting shear viscosities were
statistically indistinguishable.

\subsection{General simulation setup}
\label{sec:gen_setup}

The simulation is run in the canonical $(NVT)$ ensemble at a fixed reduced temperature $T^* = 1.5$. 
The choice of system size is a compromise between computational efficiency and simulation accuracy. As a result of their long-range nature, dipolar interactions are especially computationally demanding. 
In the present work, the fluids consist of $N = 3000$ particles. This is enough to limit finite size effects as long as $\xi \gtrsim 0.1$~\cite{Pousaneh2}.
When $\xi \gtrsim 0.35$, Ref.~\cite{Pousaneh2} observed that the dipolar interactions impact fluid structure to the extent that even pDHS theory is no longer usable.
For this reason, volume fractions $\xi > 0.35$ are not included for investigation in the present work.

All simulations are run using velocity-Verlet with a time step $\delta \tau^* = 0.001$. The fluid is equilibrated for $t^* = 2000$. During equilibration, a Nos\'e-Hoover thermostat is applied with a damping time of $\tau^*_\mathrm{damp} = 100\delta \tau^* = 0.1$.

After equilibration, the numerical viscosity is calculated using a non-equilibrium method developed by M\"uller-Plathe~\cite{Muller-Plathe1999}. The thermostat is switched to canonically sampled velocity re-scaling~\cite{Bussi2007}, with the damping time increased to $\tau^*_\mathrm{damp} = 10000\delta \tau^* = 10$. 
Although the Nos\'e-Hoover thermostat does not generally strongly impact the dynamic properties of a simulated fluid~\cite{Basconi2013}, it was previously shown in Ref.~\cite{Fjeldstad2023} that the Nos\'e-Hoover thermostat and M\"uller-Plathe method are not compatible. Canonically sampled velocity re-scaling with a long damping time does not suffer from this issue~\cite{Fjeldstad2023}.

The M\"uller-Plathe method relies on non-physical momentum swapping between particles to create a velocity gradient. The approach requires using a rectangular simulation box in order to work. 
In the present work, we set the simulation box height to three times the length of the base.
The box is divided into 100 chunks in the elongated direction and momentum is swapped between particles in the chunks at the edges and in the center, giving rise to a velocity gradient.
Since momentum is also swapped between particles with different masses, energy is not strictly conserved during swapping in our asymmetric mixtures. However, 
the rate of energy change due to swaps is much lower than the rate at which the thermostat extracts or inserts energy. We have verified that there is no significant change in the temperature after swapping begins.

The shear rate depends on how frequently a momentum swap is made, the system size and box height, and the size of the chunks being sampled. 
Assuming the fluid behavior is Newtonian, the calculated shear viscosity will not be sensitive to the exact swapping time chosen. 
In the present work, momentum swapping happens every 500 time steps, which is 0.5 in reduced time units. We have verified that the resulting velocity profiles are well-behaved and linear.
Furthermore, Ref.~\cite{Fjeldstad2023} showed that the same general setup with the same swapping rate produced a Newtonian response for the 1000 dipolar PHS particles considered there. 
As our simulations are significantly larger, with the same momentum swapping rate and thus a lower momentum flux, it is reasonable to assume that our simulations are behaving in a Newtonian manner as well.

\section{Results}
\label{sec:results}

In this section, we compare the theoretical results derived in Sec.~\ref{sec:Theory} to numerical data for the shear viscosity of simulated fluids as described in Sec.~\ref{sec:SimSetUp}.
We first focus on the pure fluids and their parameters, and then move on to the mixture parameters and binary mixtures.

\subsection{The shear viscosity of pure DHS fluids}
\label{sec:pure_DHS}

\begin{figure}
    \centering
    \subfloat[]{
        \includegraphics[width=0.45\textwidth]{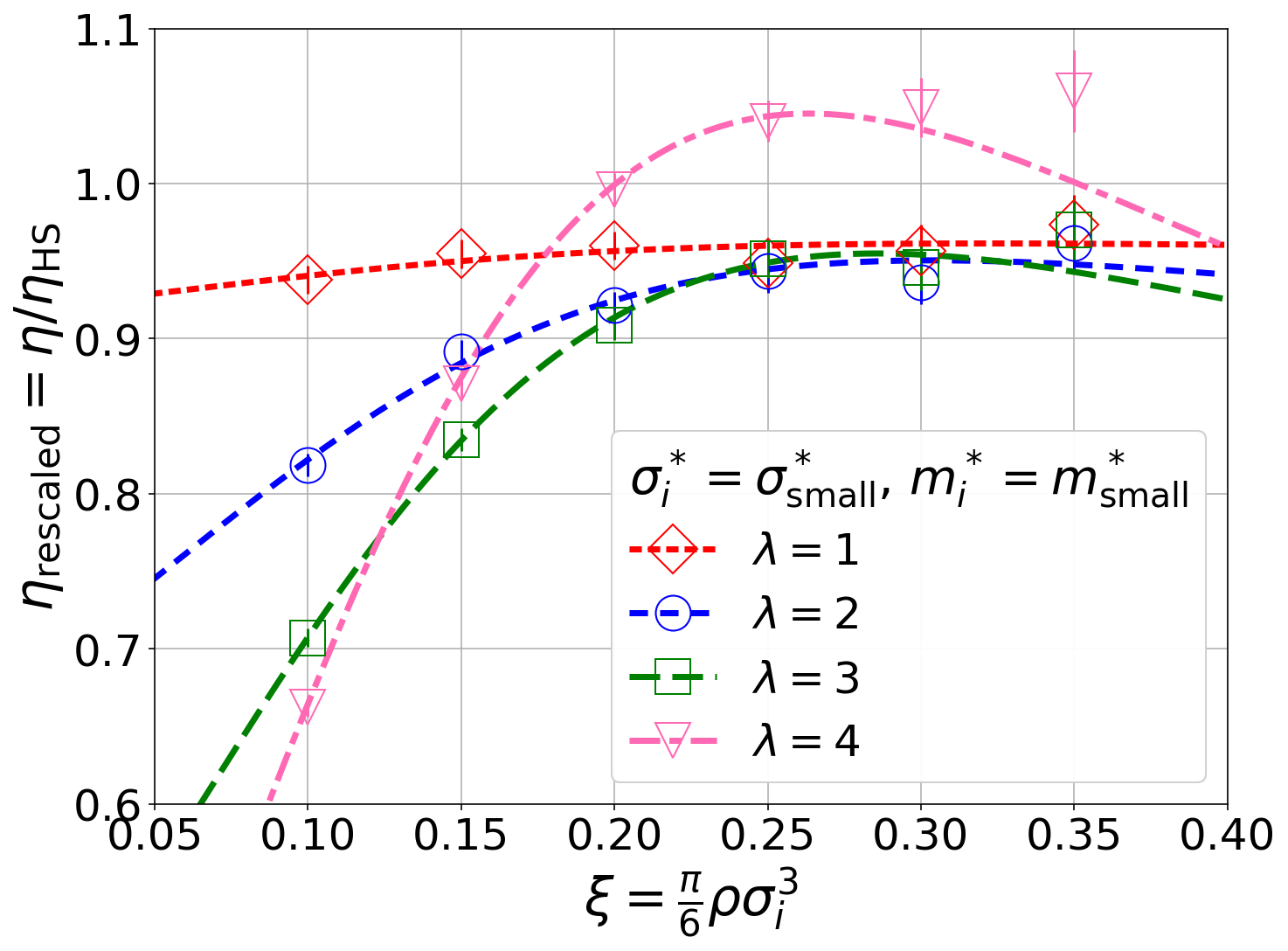}
        \label{subfig:pure_A}}
        \newline
    \subfloat[]{
        \includegraphics[width=0.45\textwidth]{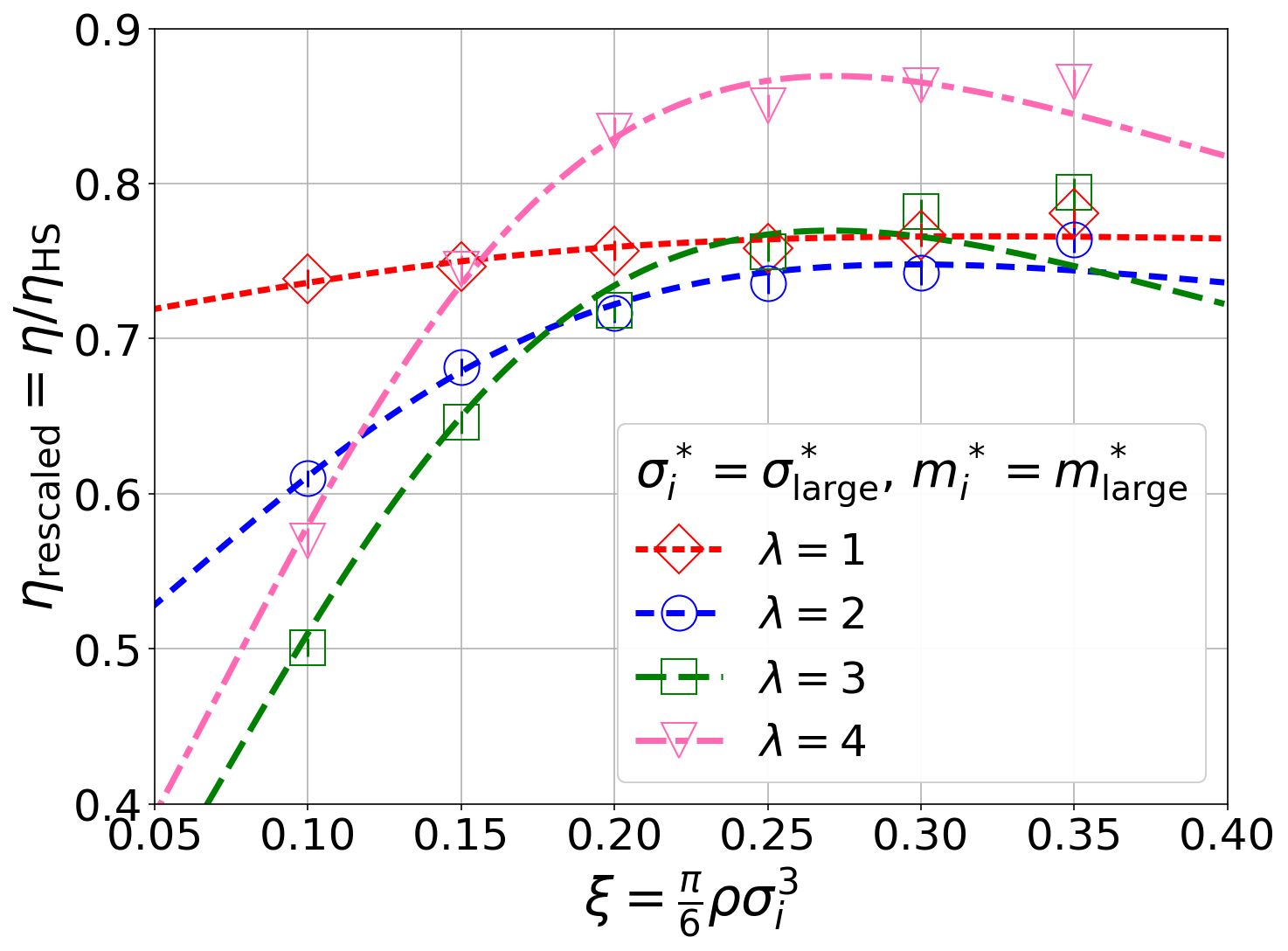}
        \label{subfig:pure_B}}
    \caption{Rescaled shear viscosity from simulations for pure DHS fluids with $\lambda = 1$, $2$, $3$ and $4$. The rescaling uses the Enskog expression for a pure hard-sphere fluid at the same volume fraction with $\sigma_i^* = 1$, mass $m_i^* = 1$, and reduced temperature $T^* = 1.5$. The predictions are made using bmDHS theory for fluids with small particles ($\sigma_i^* = 1$, $m_i^* = 1$) \protect\subref{subfig:pure_A} and large particle fluids ($\sigma_i^* = 1.5$, $m_i^* = 3.375$) \protect\subref{subfig:pure_B}. The associated fit parameters can be found in tables~\ref{tab:fit_parameters_pure}.}
    \label{fig:Viscosity_pure_fluids}
\end{figure}

We begin by focusing on the limit of pure fluids, comparing bmDHS theory to pDHS theory, and extract effective parameters that we later use for predicting the composition dependence of the mixture viscosity. 
The effective parameters $\lambda^\mathrm{eff}_{ii}$ and $\Omega^\mathrm{eff}_{ii}$
are determined by fitting Eq.~(\ref{eq:Enskog}) to data for the shear viscosity of simulated pure DHS fluids using $\chi_i = \chi^\mathrm{DHS}_{ii}|_{\lambda_{ii} = \lambda^\mathrm{eff}_{ii}, x_i=1}$.
The results for small particle fluids ($\sigma^*_i = 1$ and $m^*_i = 1$) are shown in Fig.~\ref{fig:Viscosity_pure_fluids}\subref{subfig:pure_A}. The results for large particle fluids ($\sigma^*_i = 1.5$ and $m^*_i = 3.375$) are shown in Fig.~\ref{fig:Viscosity_pure_fluids}\subref{subfig:pure_B}.
The moment of inertia is proportional to $m^*_i\sigma^{*2}_i$ (see section~\ref{sec:sim_params} for details), so that
the larger particle species has a moment of inertia $1.5^5\approx7.6$ times greater than that of the smaller particle species.

The theory is fit to simulation data from the full range of densities investigated.
Values for the fit parameters $\lambda^\mathrm{eff}_{ii}$ and $\Omega^\mathrm{bmDHS}_{ii}$, along with the associated values calculated for $\Omega^\mathrm{eff}_{ii}$, are given in table~\ref{tab:fit_parameters_pure}. 
Note that all shear viscosity data has been rescaled using the Enskog expression for a pure hard-sphere fluid at the same volume fraction with $\sigma_i^* = 1$, mass $m_i^* = 1$, and reduced temperature $T^* = 1.5$. This is done for visual clarity, as the actual viscosity changes by more than an order of magnitude over this range. The same rescaling is applied in all figures showing shear viscosity data.

\begin{table}[t!]
\centering
\begin{tabular}{ p{0.18\linewidth} c c c p{0.1\linewidth} c c c}
\hline
\hline
\multicolumn{8}{c}{\textbf{Fit parameters using bmDHS theory}}\\
\hline
 & \multicolumn{3}{c}{$\sigma_i^* = 1$, $m_i^* = 1,$} & & \multicolumn{3}{c}{$\sigma_i^* = 1.5$, $m_i^* = 3.375,$} \\
 & \multicolumn{3}{c}{$I^*_i = 0.0099856 m^*_i \sigma^*_i $} & & \multicolumn{3}{c}{$I_i^* = 0.07582815 m_i^* \sigma_i^* $ } \\
 & $\lambda^\mathrm{eff}_{ii}$ & $\Omega^\mathrm{bmDHS}_{ii}$ & $\Omega^\mathrm{eff}_{ii}$ & & $\lambda^\mathrm{eff}_{ii}$ & $\Omega^\mathrm{bmDHS}_{ii}$ & $\Omega^\mathrm{eff}_{ii}$ \\
\hline
$\lambda_i = 1$ & 0.34 & 1.05 & 1.09 & & 0.47 & 1.08 & 1.16 \\
$\lambda_i = 2$ & 0.96 & 1.11 & 1.48 & & 1.18 & 1.18 & 1.71 \\
$\lambda_i = 3$ & 1.55 & 1.19 & 2.40 & & 2.03 & 1.31 & 3.06 \\
$\lambda_i = 4$ & 2.26 & 1.25 & 4.94 & & 2.00 & 1.15 & 2.49 \\
\hline
\end{tabular}
\caption{Fit parameters $\lambda^\mathrm{eff}_{ii}$ and $\Omega^\mathrm{bmDHS}_{ii}$ obtained by fitting Eq.~(\ref{eq:Enskog}) to simulation data using bmDHS theory for expressions for $\chi_i = \chi^\mathrm{DHS}_{ii}$ with $x_j = 0$. The table also includes the values calculated for $\Omega^\mathrm{eff}_{ii}$ by reordering Eq.~(\ref{eq:Omega_bmDHS}).}
\label{tab:fit_parameters_pure}
\end{table}

The fits presented in Fig.~\ref{fig:Viscosity_pure_fluids} are good in all cases, both for small (Fig.~\ref{fig:Viscosity_pure_fluids}\subref{subfig:pure_A}) and large (Fig.~\ref{fig:Viscosity_pure_fluids}\subref{subfig:pure_B}) particle fluids for all densities $\xi \lesssim 0.35$ depending on the strength of the dipolar coupling. As noted in Sec.~\ref{sec:gen_setup}, structural changes occur in DHS fluids above this density, leading to long-range order~\cite{Weis2006, Pousaneh2} not accounted for by Chapman-Enskog theory. Qualitatively, the fit using bmDHS theory is very similar to that obtained by Pousaneh and de Wijn using pDHS theory, although quantitatively, both the exact expression used and the simulation setup are slightly different, and so the resulting effective parameters are also different. 
We conclude that, in the pure limit, the re-summed expression for $\chi^\mathrm{DHS}_{ii}$ (Eq.~\ref{eq:chi_as_a_sum_BMCSL_resum}), which is a crucial ingredient in bmDHS theory, is able to reproduce the viscosity with an acceptable degree of accuracy.

In the case of the large particle fluids, a very strong dipolar coupling strength appears to be associated with a decrease in both $\lambda^\mathrm{eff}_{ii}$ and $\Omega^\mathrm{eff}_{ii}$ as functions of the true dipolar coupling constant $\lambda_{i}$ (see table~\ref{tab:fit_parameters_pure}).
We suspect that this is related to the higher moment of inertia of the larger particles, as it impacts the rotational degrees of freedom, which couple more strongly when large dipole moments are present. This means the collision is no longer even approximately elastic, and momentum exchange during the collision is different.

\subsection{Empirical expression for \texorpdfstring{$\Omega(\lambda)$}{Om(L)}}

\begin{figure}
    \centering
    \includegraphics[width=0.45\textwidth]{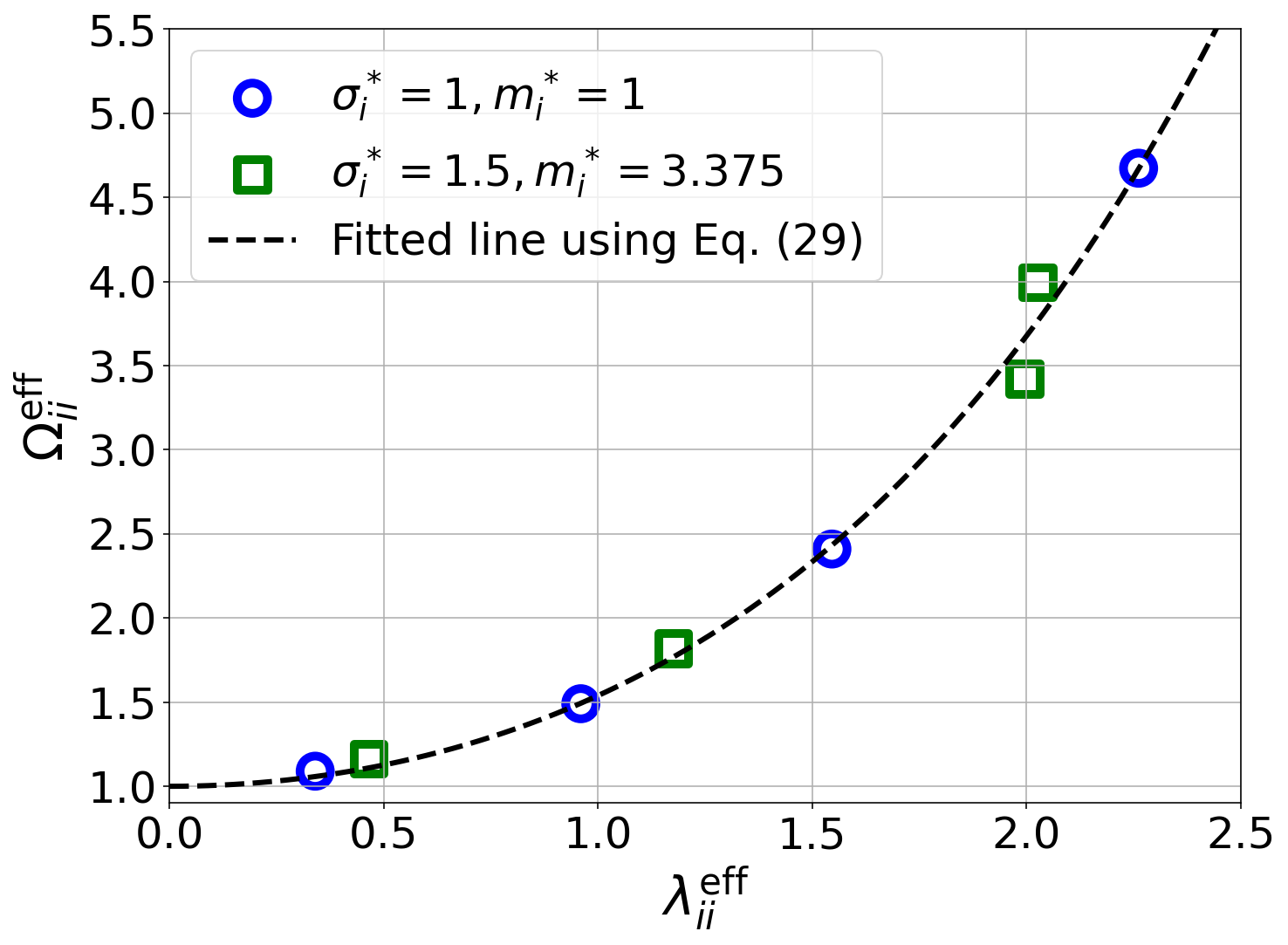}
    \caption{The values calculated for $\Omega^\mathrm{eff}_{ii}$ are plotted as a function of $\lambda^\mathrm{eff}_{ii}$ for both the small particle (circles) and large particle (squares) pure fluids ($\sigma^*_i = 1$ and $m^*_i = 1$, and $\sigma^*_i = 1.5$ and $m^*_i = 3.375$ respectively). The figure also includes the curve (dashed line) which results from fitting Eq.~(\ref{eq:Omega_empirical}) to the pure fluid data.}
    \label{fig:Omega_star_vs_L_eff}
\end{figure}

In order to obtain $\Omega^\mathrm{eff}_{ii}$, we must first investigate how $\Omega^\mathrm{eff}_{ii}$ is related to $\lambda^\mathrm{eff}_{ii}$, as discussed in Sec.~\ref{sec:1-2-params}.
Figure~\ref{fig:Omega_star_vs_L_eff} shows $\Omega^\mathrm{eff}_{ii}$ as a function of $\lambda^\mathrm{eff}_{ii}$ both when the fluid particles are small ($\sigma^*_i = 1$ and $m^*_i = 1$) and large ($\sigma^*_i = 1.5$ and $m^*_i = 3.375$). 
This plot suggests that indeed, $\Omega^\mathrm{eff}_{ii}$ has a universal dependence on the effective dipolar coupling constant, regardless of other parameters,
at least for the dipolar coupling constants and particle sizes included in the present investigation.

We therefore suggest an empirical expression for $\Omega(\lambda)$ given as an even power series in $\lambda$ which, for the present purpose, is truncated beyond the fourth power.
The choice of an even power series in $\lambda$ is loosely motivated by the observation that
any terms containing odd powers of one dipole moment or the other will vanish due to orientational averaging.
Since the coupling constant is the product of the two dipole moments, any odd-powered terms thus vanish.
Similarly, in the cluster expansion of $\chi^\mathrm{DHS}_{ii}$, only even powers in the dipolar coupling constant contribute to the partial PDF in the low density limit and the odd powers vanish due to orientational averaging when no external field is present~\cite{Elfimova2010}.
In addition, in order to show the correct behavior in the weak coupling limit, the expression must maintain consistency with hard-sphere theory when $\lambda$ goes to zero, and recover the correct hard-sphere value $\Omega^\mathrm{eff}_{ii} = 1$.
The final expression is
\begin{equation}
\label{eq:Omega_empirical}
    \Omega(\lambda) = 1 + a_1 {\lambda}^2 + a_2 {\lambda}^4~.
\end{equation}
The coefficients are obtained by fitting the expression to the data for pure fluids given in table~\ref{tab:fit_parameters_pure},
giving $a_1 = 0.50$ and $a_2 = 0.04$.
The resulting fit is also included in Fig.~\ref{fig:Omega_star_vs_L_eff} (dashed line). As is clear from the figure, the fit is excellent for the range of parameters investigated. 
Crucially, this now allows us to estimate $\Omega^\mathrm{eff}_{ij}$ for any combination of dipolar species, without any need for further fitting to viscosity data.
Combining Eq.~(\ref{eq:Omega_empirical}) with Eq.~(\ref{eq:omega_eff_12}) provides a path for estimating $\Omega^\mathrm{eff}_{12}$. 

Although Eq.~(\ref{eq:Omega_empirical}) does appear to fit the data well, it is likely that there are other functional forms for $\Omega(\lambda)$ that would provide an equally good fit. As the expression is not intended for use outside of the range of dipolar coupling constants for which it was fitted, any expression that appears to smoothly interpolate between the included data points is likely to provide similar predictions when combined with Eq.~(\ref{eq:omega_eff_12}) for estimating $\Omega^\mathrm{eff}_{12}$. 
A more detailed investigation is beyond the scope of this work, but would clearly be useful for applying our theory.

\subsection{Composition-dependent shear viscosity}

We now turn our attention to the binary mixtures, and compare the predictions to the simulation results.
For additional comparison, one might consider alternative approaches to predicting the shear viscosity. The standard alternative approach for application of Enskog theories, such as the Enskog-Thorne expression, to non-hard-sphere systems is to assign an effective hard-sphere diameter and apply the relevant hard-sphere expressions. This approach has however already been demonstrated to be extremely unsuitable for pure DHS fluids~\cite{Pousaneh2}.

\subsubsection{The case when \texorpdfstring{$\sigma^*_1 = \sigma^*_2$}{sigma1 = sigma2}}

\begin{figure*}
    \centering
    \subfloat[]{
        \includegraphics[width=0.49\textwidth]{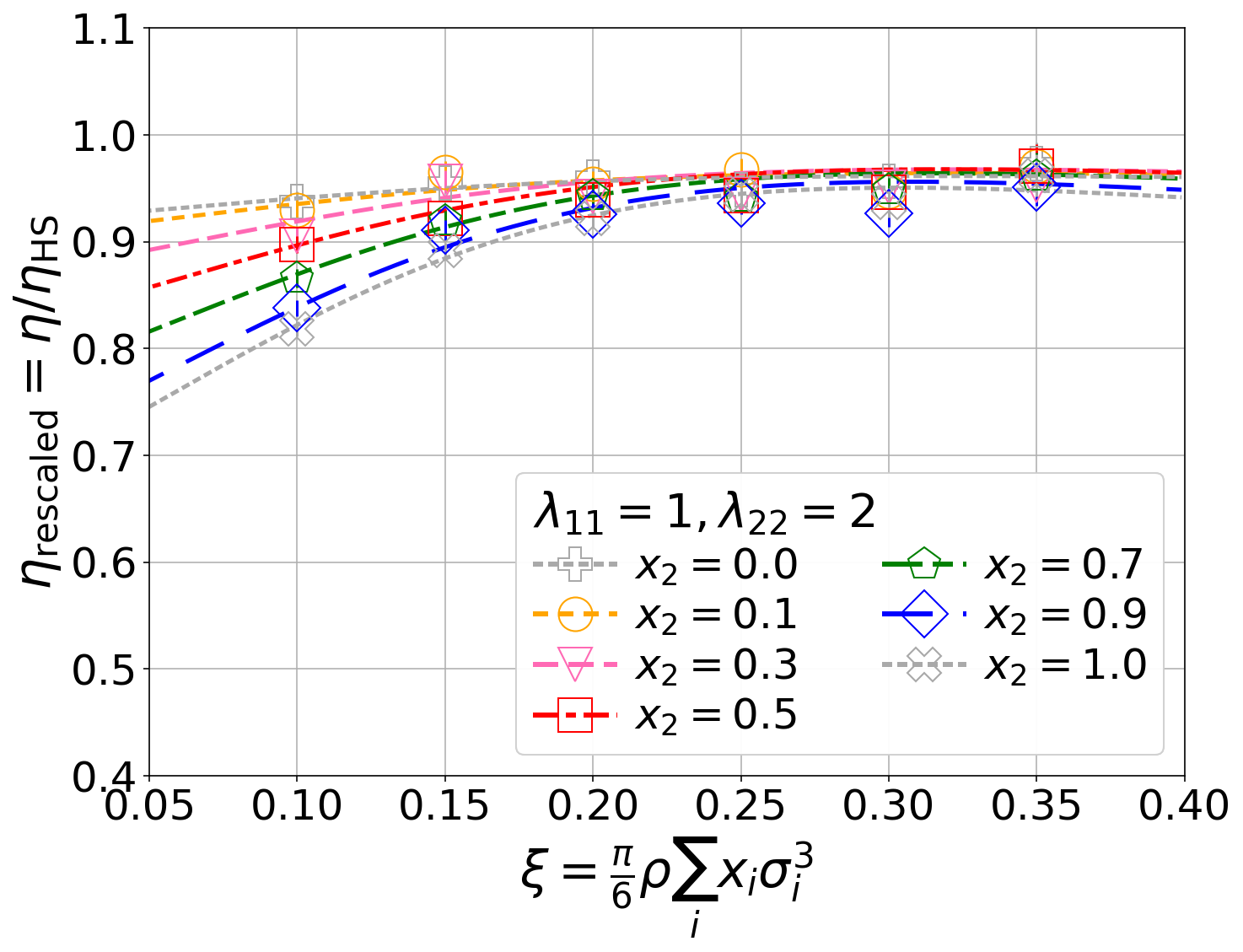}
        \label{subfig:A}}
    \subfloat[]{
        \includegraphics[width=0.49\textwidth]{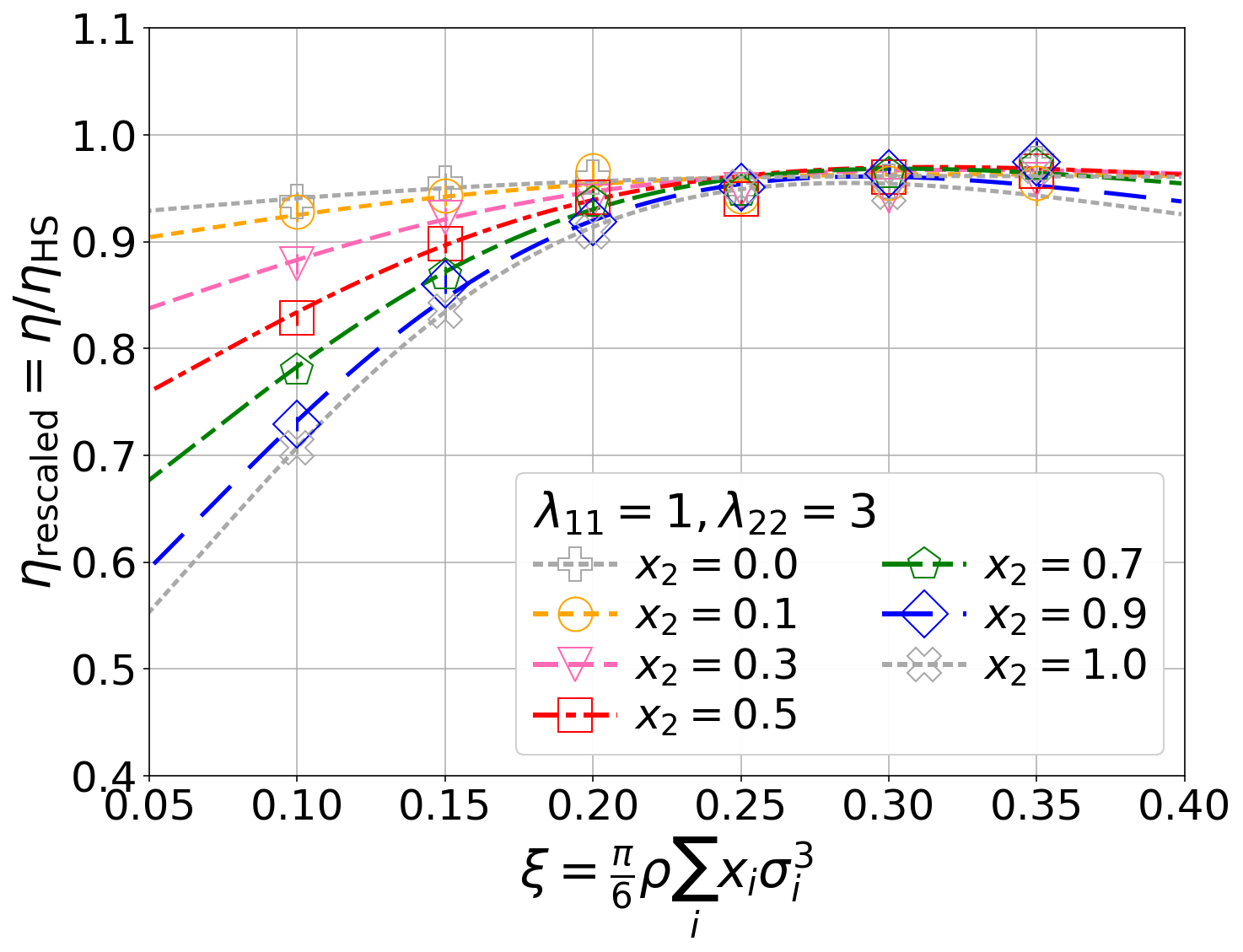}
        \label{subfig:B}}
        \newline
    \subfloat[]{
        \includegraphics[width=0.49\textwidth]{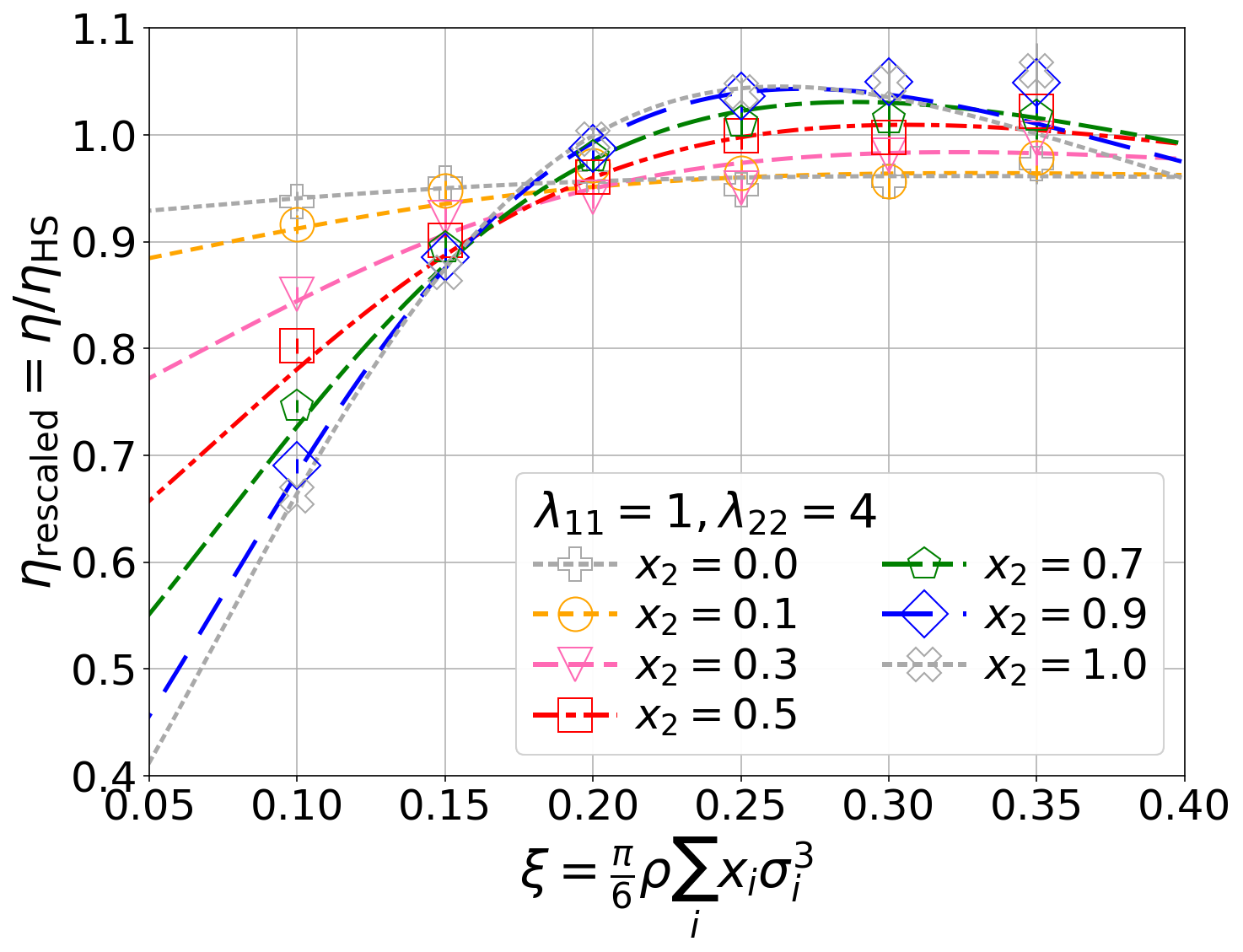}
        \label{subfig:C}}
    \subfloat[]{
        \includegraphics[width=0.49\textwidth]{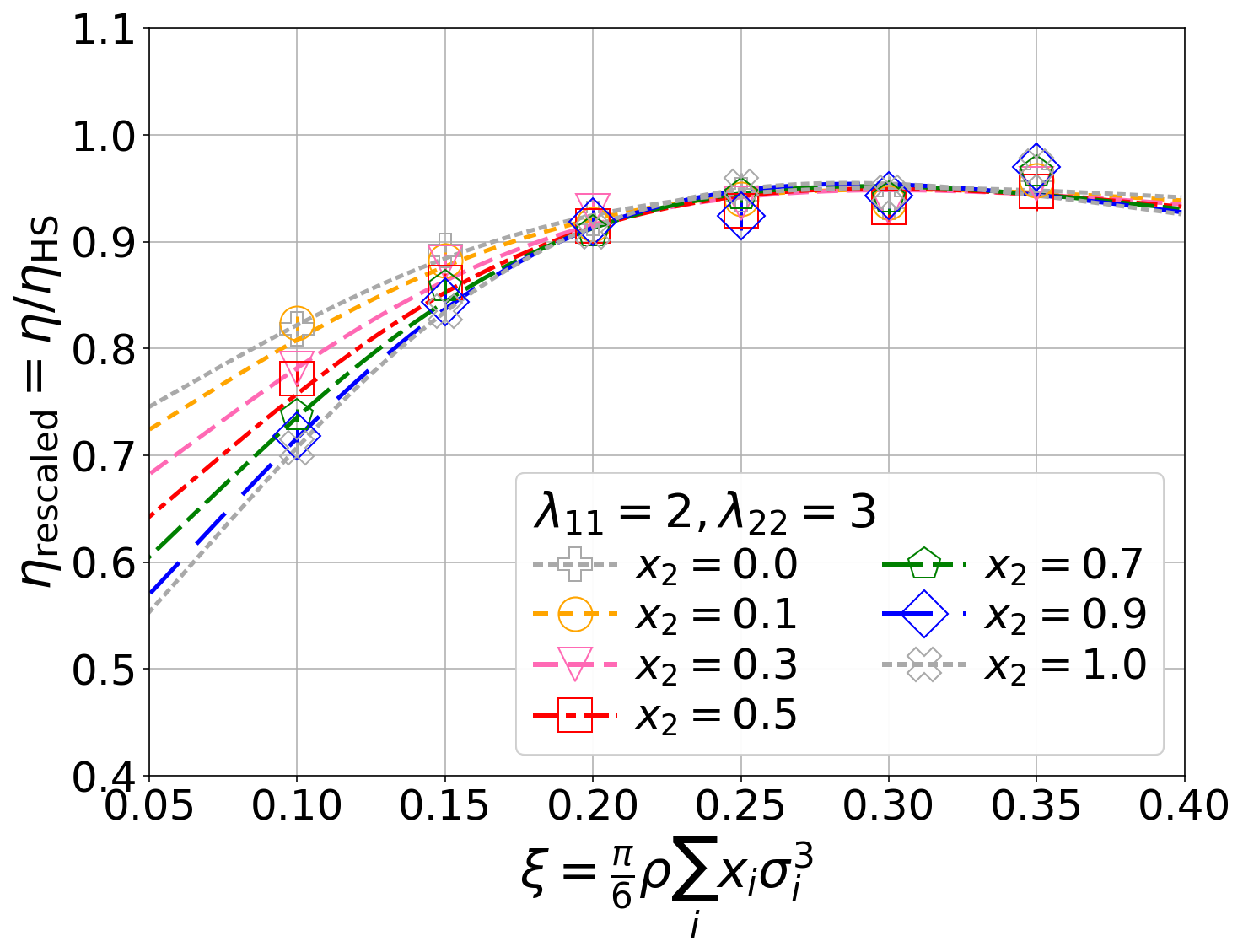}
        \label{subfig:D}}
        \newline
    \subfloat[]{
        \includegraphics[width=0.49\textwidth]{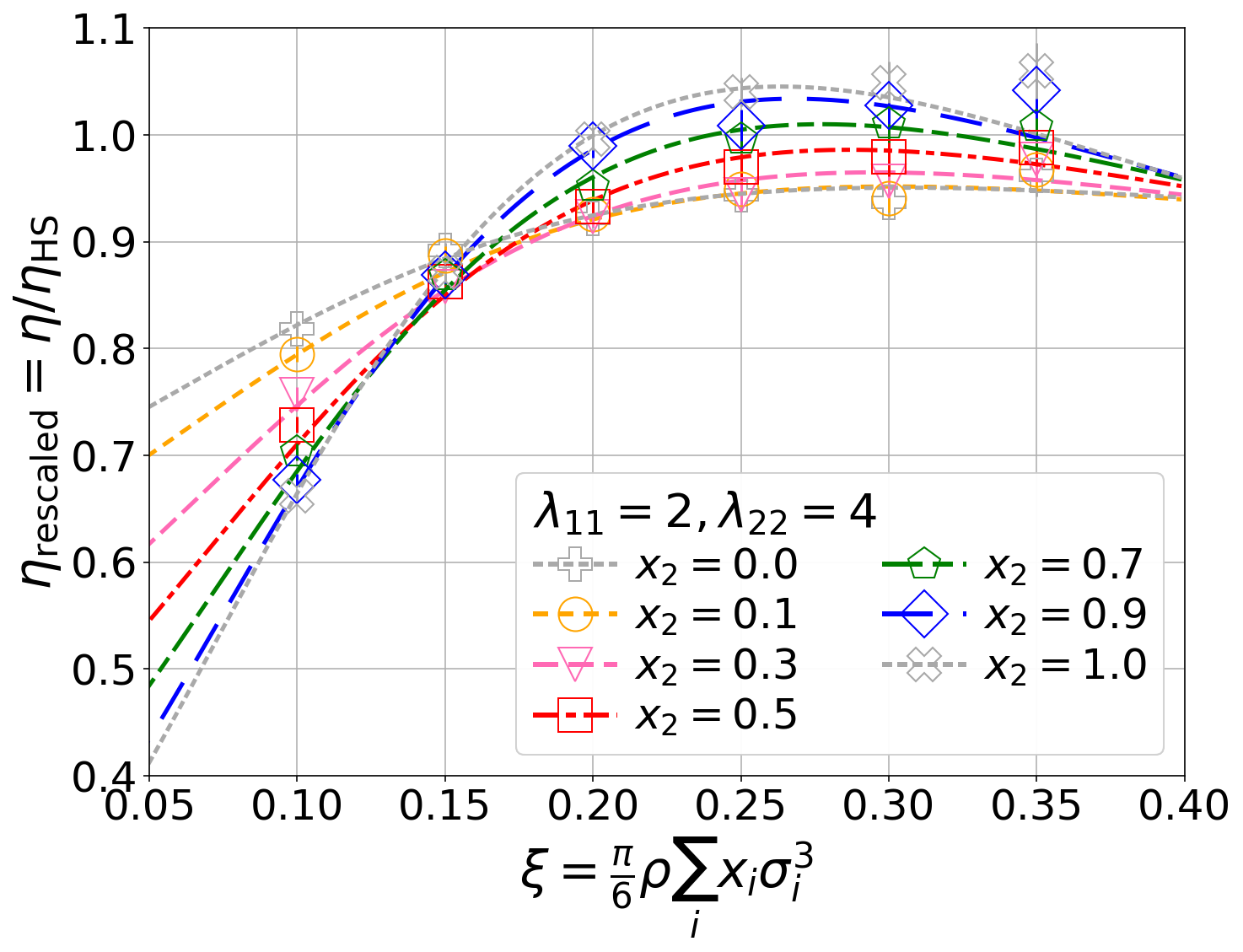}
        \label{subfig:E}}
    \subfloat[]{
        \includegraphics[width=0.49\textwidth]{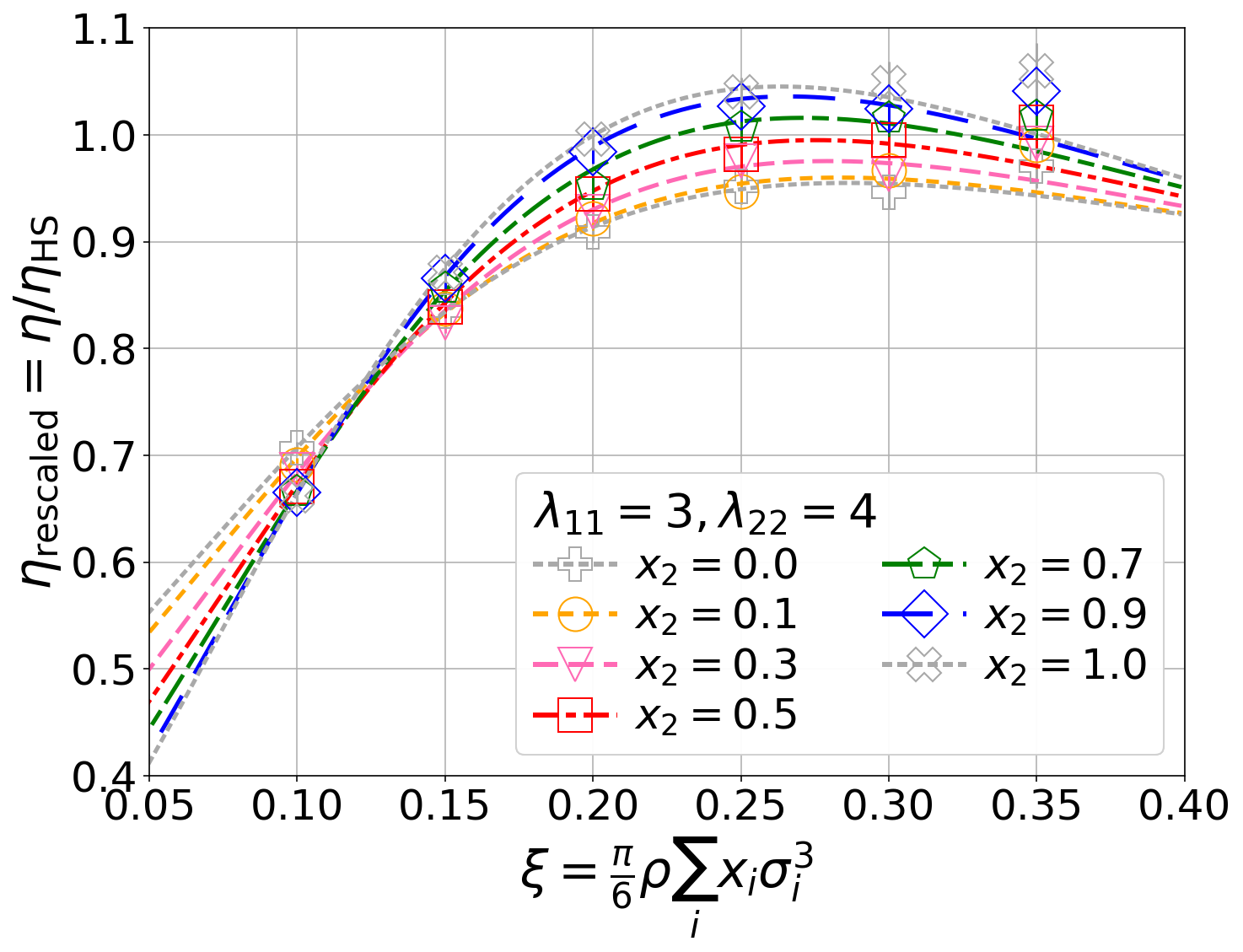}
        \label{subfig:F}}
    \caption{Rescaled shear viscosity from simulations of mixtures where both components have equally sized cores and (a) $(\lambda_{11}, \lambda_{22}) = (1, 2)$, (b) $(\lambda_{11}, \lambda_{22}) = (1, 3)$, (c) $(\lambda_{11}, \lambda_{22}) = (1, 4)$, (d) $(\lambda_{11}, \lambda_{22}) = (2, 3)$, (e) $(\lambda_{11}, \lambda_{22}) = (2, 4)$, and (f) $(\lambda_{11}, \lambda_{22}) = (3, 4)$, along with the predicted shear viscosity of the binary DHS fluid mixtures using bmDHS theory. The rescaling uses the Enskog expression for a pure hard-sphere fluid at the same volume fraction with $\sigma_i^* = 1$, mass $m_i^* = 1$, and reduced temperature $T^* = 1.5$.}
    \label{fig:Viscosity_DHS_theory}
\end{figure*}

Figure~\ref{fig:Viscosity_DHS_theory} shows the shear viscosity of the simulated binary DHS fluids consisting of a mixture of two species with equally-sized cores but different dipole moments. The figure also includes the predictions made using bmDHS theory. The figure shows that bmDHS provides accurate predictions for the shear viscosity of the simulated mixture for all combinations of $\lambda_{11}$ and $\lambda_{22}$ considered and for all $\xi \leq 0.3$.
For densities $\xi > 0.3$, bmDHS theory begins to fail, especially when very strong dipoles are present, i.e. when $\lambda_{22} = 4$, corresponding to Figs.~\ref{fig:Viscosity_DHS_theory}\subref{subfig:C},~\ref{fig:Viscosity_DHS_theory}\subref{subfig:E} and~\ref{fig:Viscosity_DHS_theory}\subref{subfig:F}.

\subsubsection{The case of fully asymmetric mixtures}
\label{sec:different_sizes}

\begin{figure*}
    \centering
    \subfloat[]{
        \includegraphics[width=0.49\textwidth]{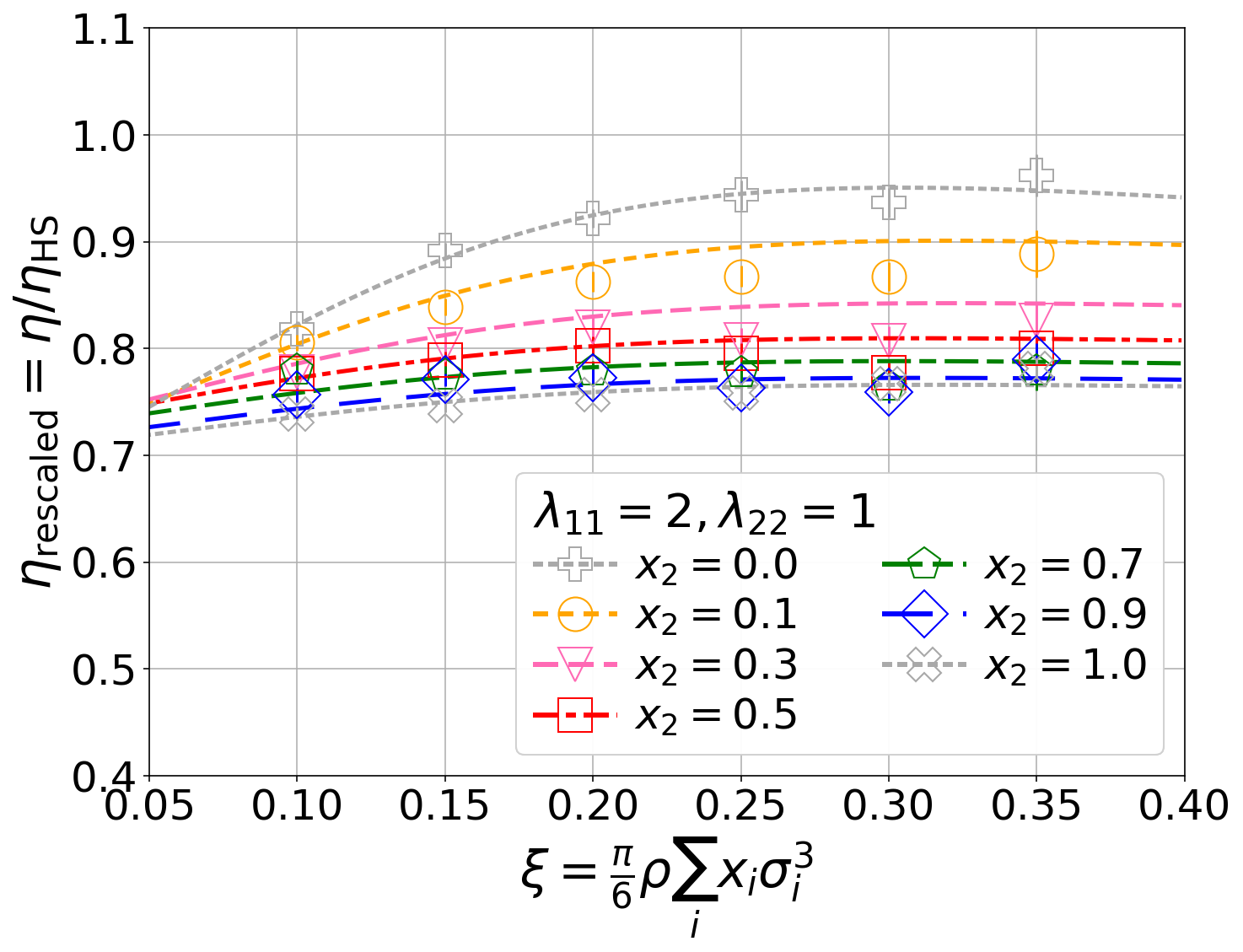}
        \label{subfig:A_different_small}}
    \subfloat[]{
        \includegraphics[width=0.49\textwidth]{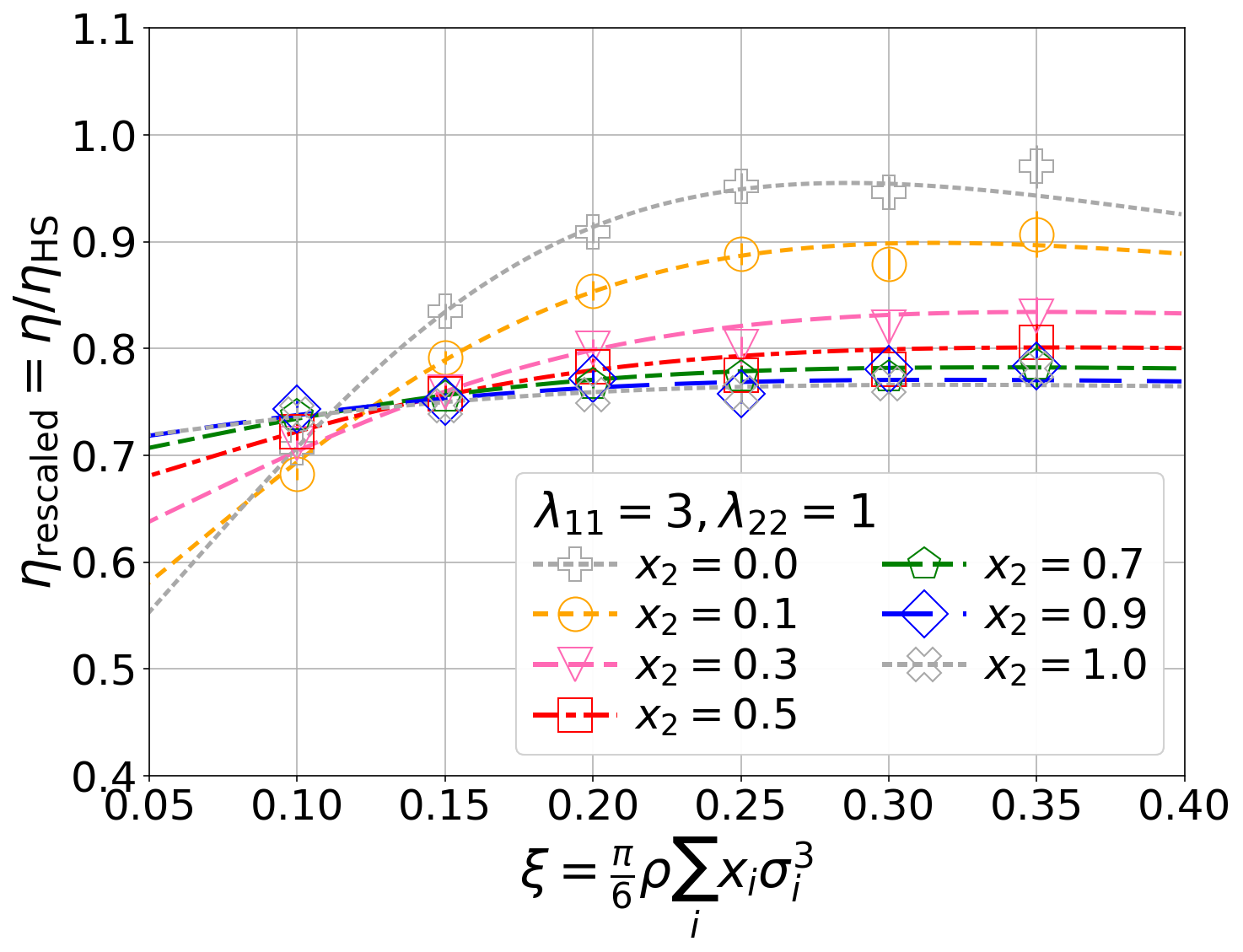}
        \label{subfig:B_different_small}}
        \newline
    \subfloat[]{
        \includegraphics[width=0.49\textwidth]{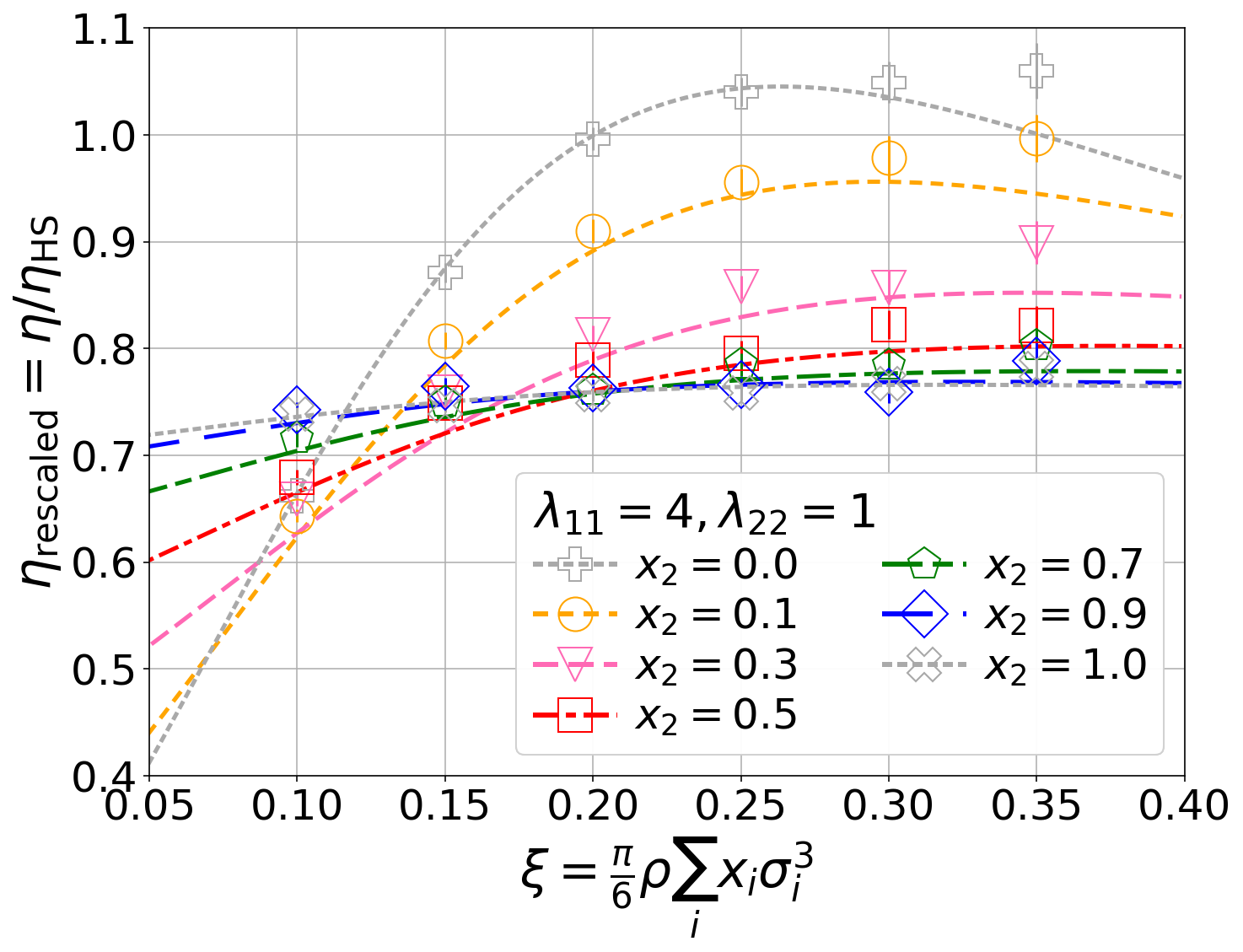}
        \label{subfig:C_different_small}}
    \subfloat[]{
        \includegraphics[width=0.49\textwidth]{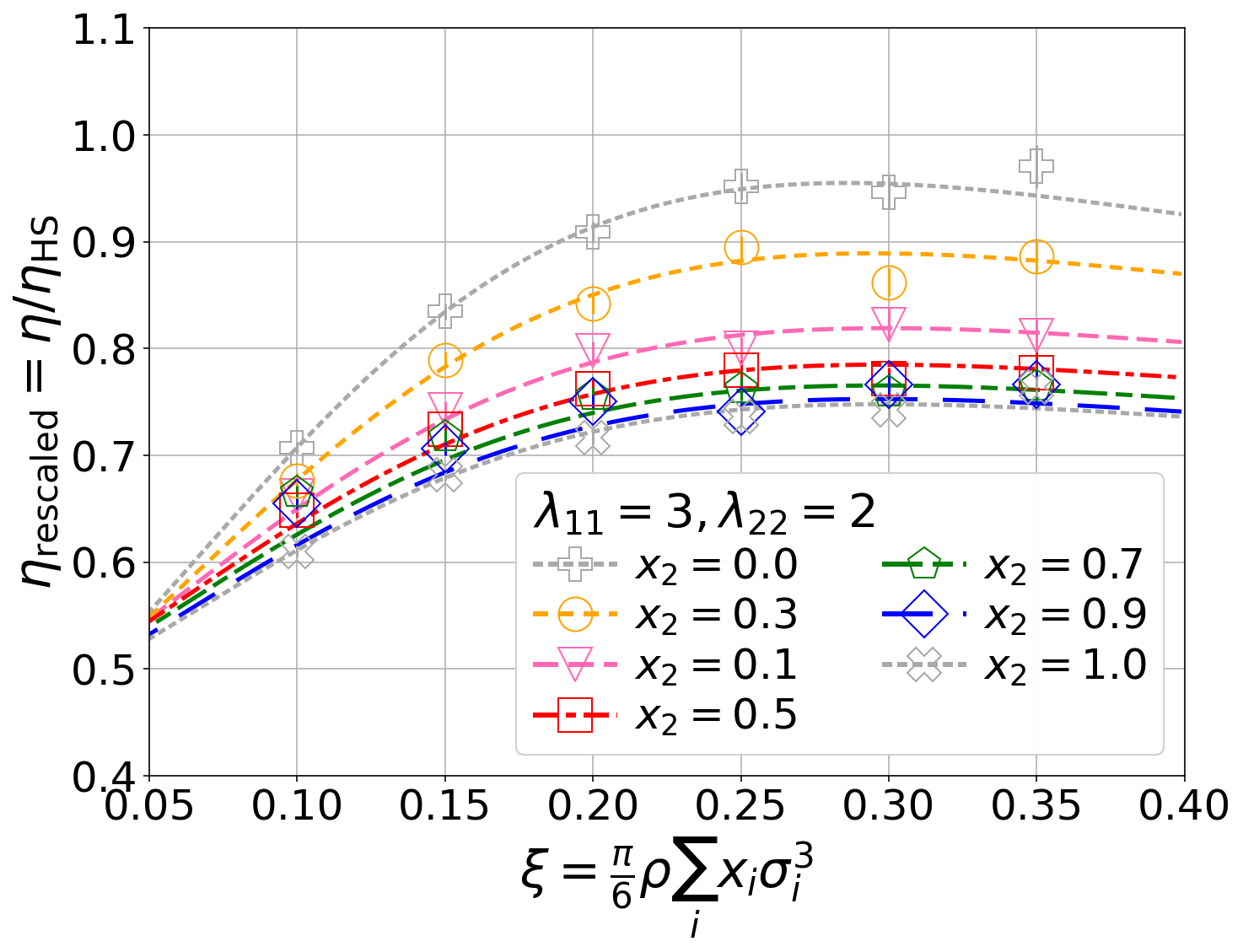}
        \label{subfig:D_different_small}}
        \newline
    \subfloat[]{
        \includegraphics[width=0.49\textwidth]{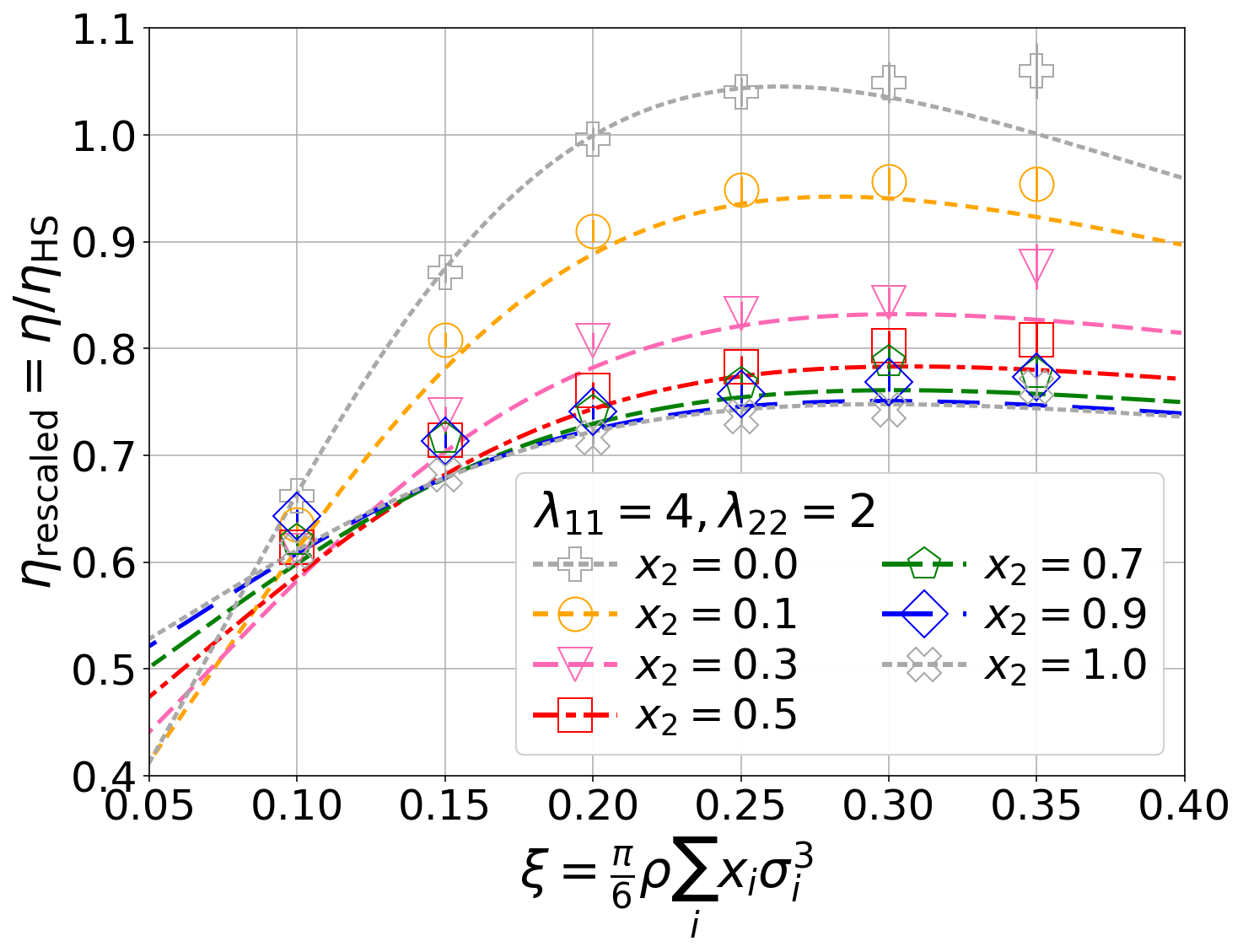}
        \label{subfig:E_different_small}}
    \subfloat[]{
        \includegraphics[width=0.49\textwidth]{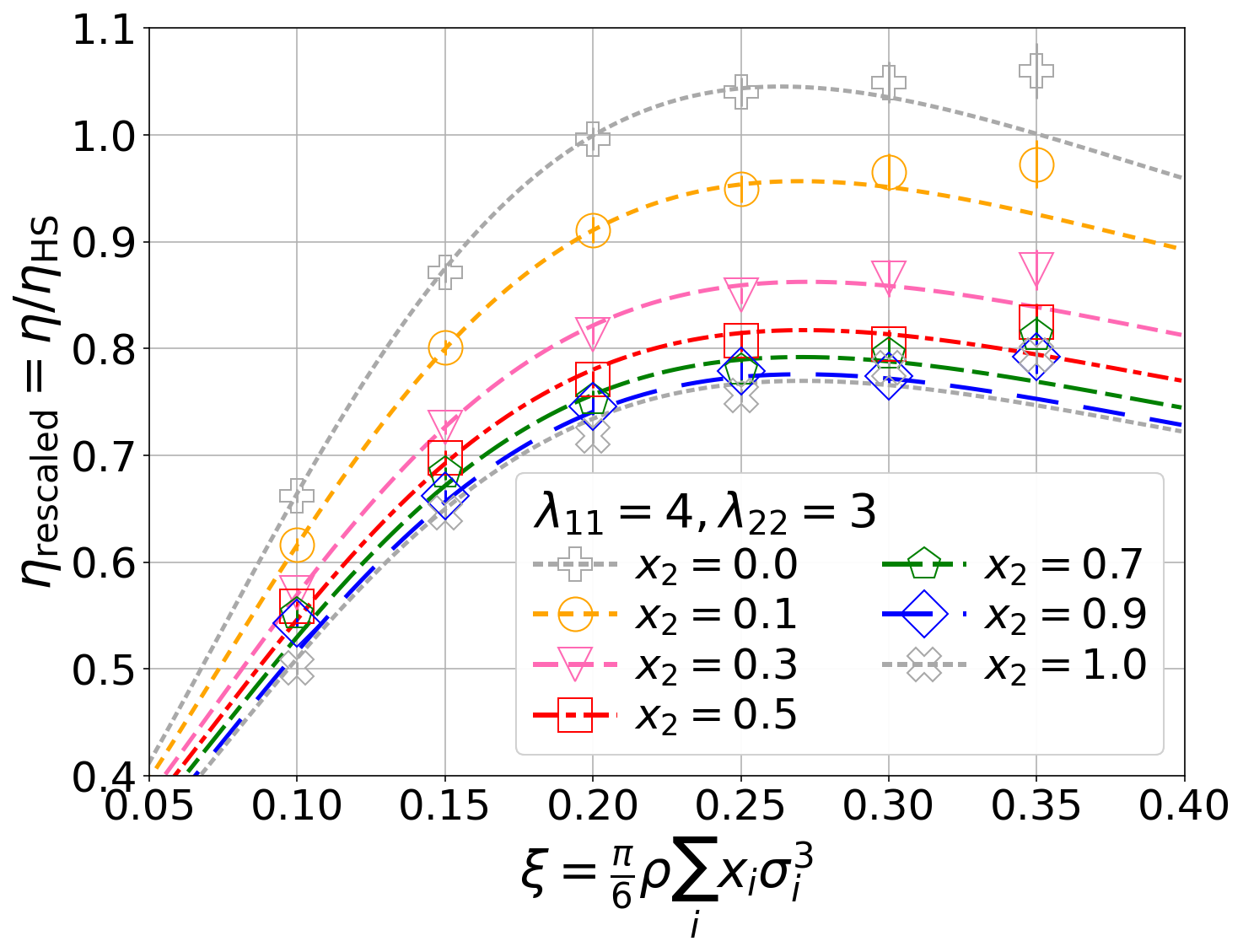}
        \label{subfig:F_different_small}}
    \caption{Rescaled shear viscosity from simulations of mixtures where the components have differently sized cores and (a) $(\lambda_{11}, \lambda_{22}) = (1, 2)$, (b) $(\lambda_{11}, \lambda_{22}) = (1, 3)$, (c) $(\lambda_{11}, \lambda_{22}) = (1, 4)$, (d) $(\lambda_{11}, \lambda_{22}) = (2, 3)$, (e) $(\lambda_{11}, \lambda_{22}) = (2, 4)$, and (f) $(\lambda_{11}, \lambda_{22}) = (3, 4)$, along with the predicted shear viscosity of the binary DHS fluid mixtures using bmDHS theory. The rescaling uses the Enskog expression for a pure hard-sphere fluid at the same volume fraction with $\sigma_i^* = 1$, mass $m_i^* = 1$, and reduced temperature $T^* = 1.5$.}
    \label{fig:Viscosity_DHS_theory_different_small}
\end{figure*}

\begin{figure*}
    \centering
    \subfloat[]{
        \includegraphics[width=0.49\textwidth]{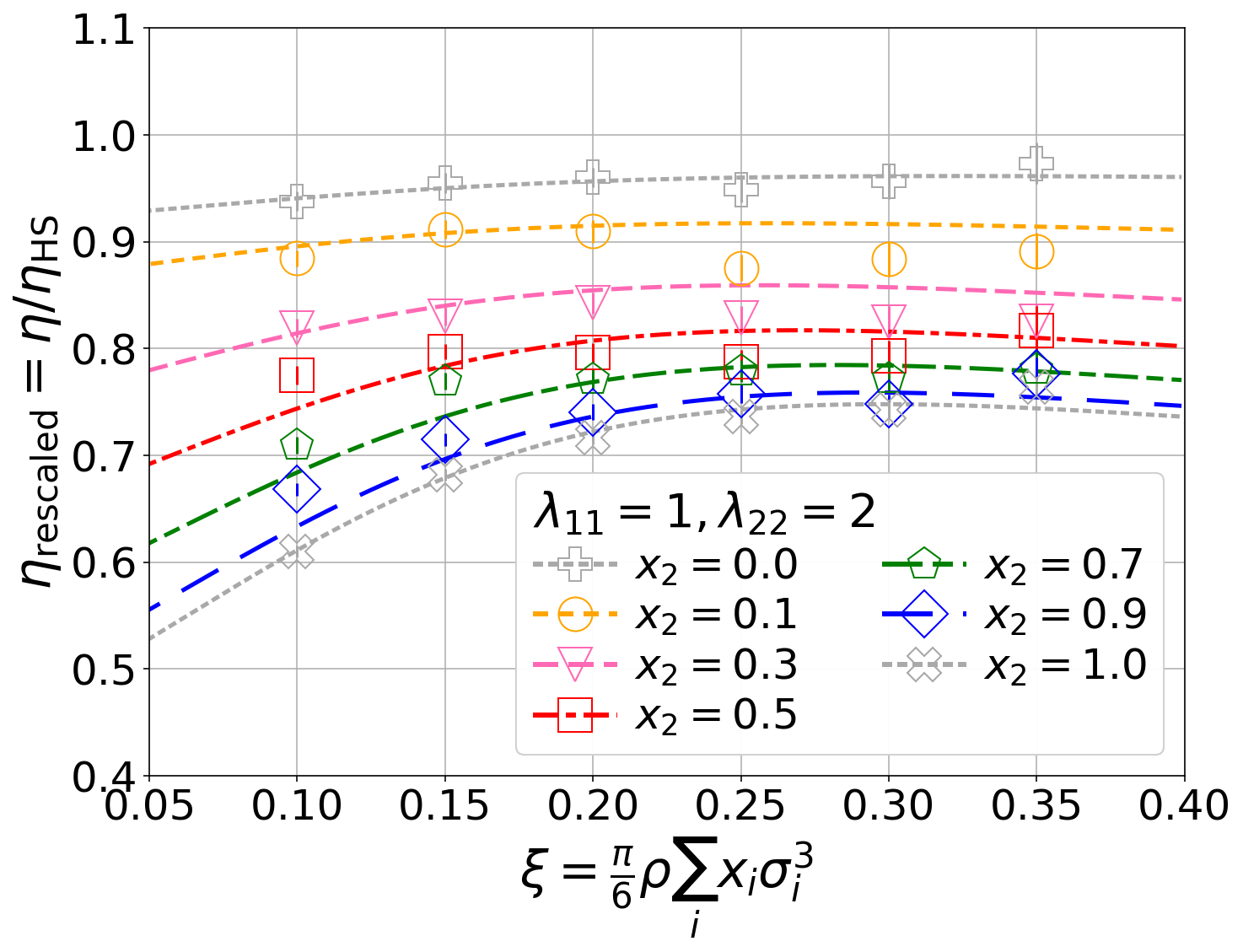}
        \label{subfig:A_different_large}}
    \subfloat[]{
        \includegraphics[width=0.49\textwidth]{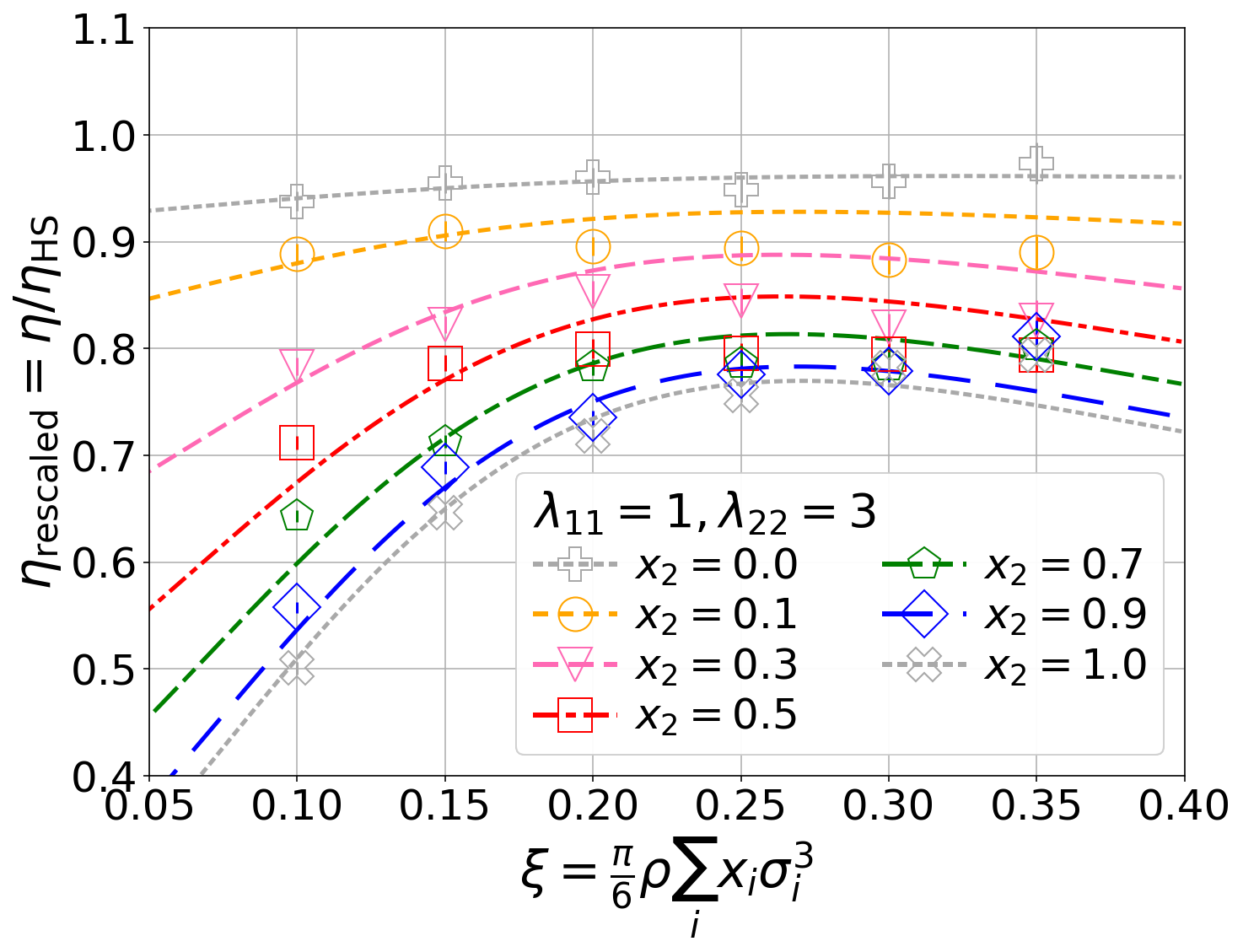}
        \label{subfig:B_different_large}}
        \newline
    \subfloat[]{
        \includegraphics[width=0.49\textwidth]{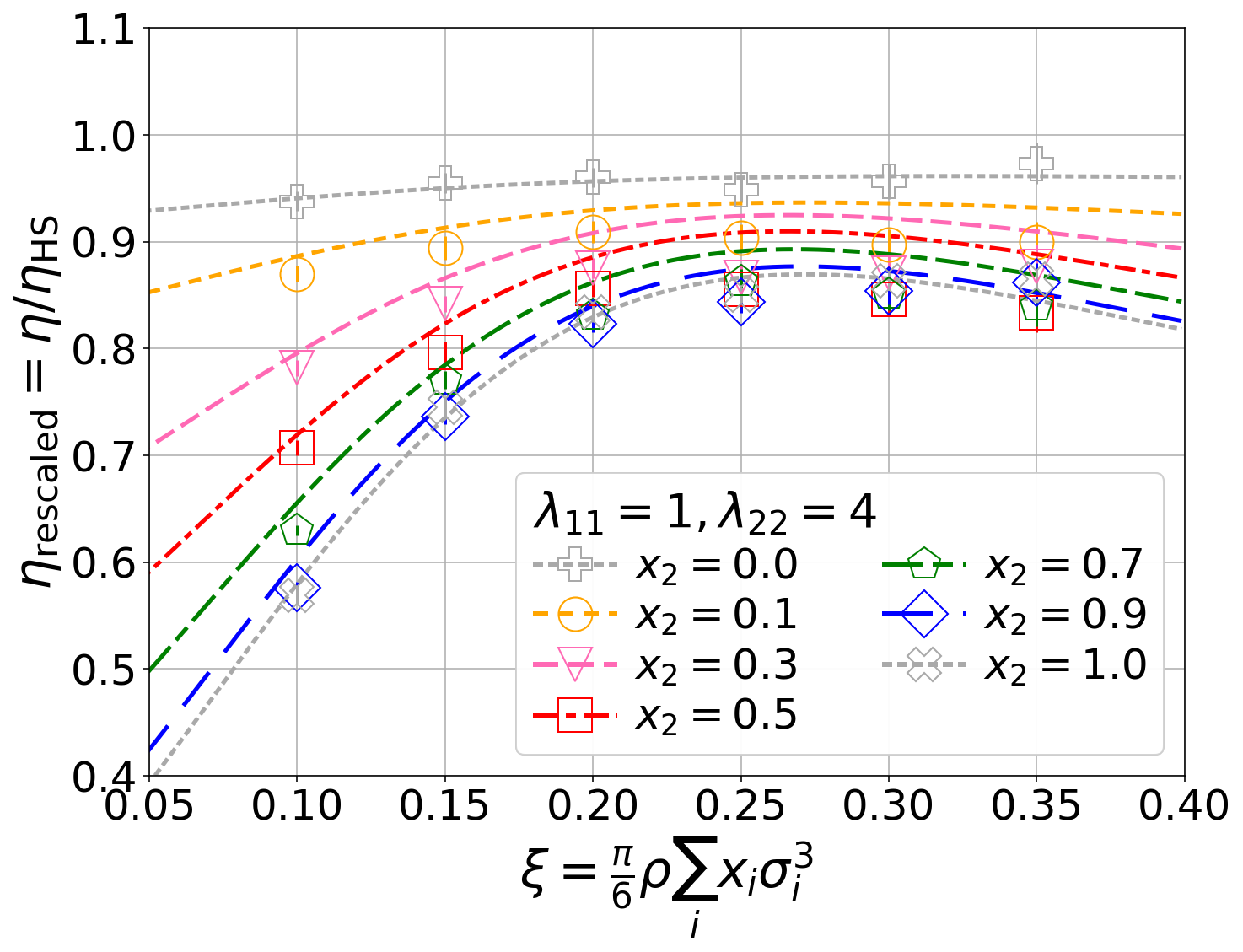}
        \label{subfig:C_different_large}}
    \subfloat[]{
        \includegraphics[width=0.49\textwidth]{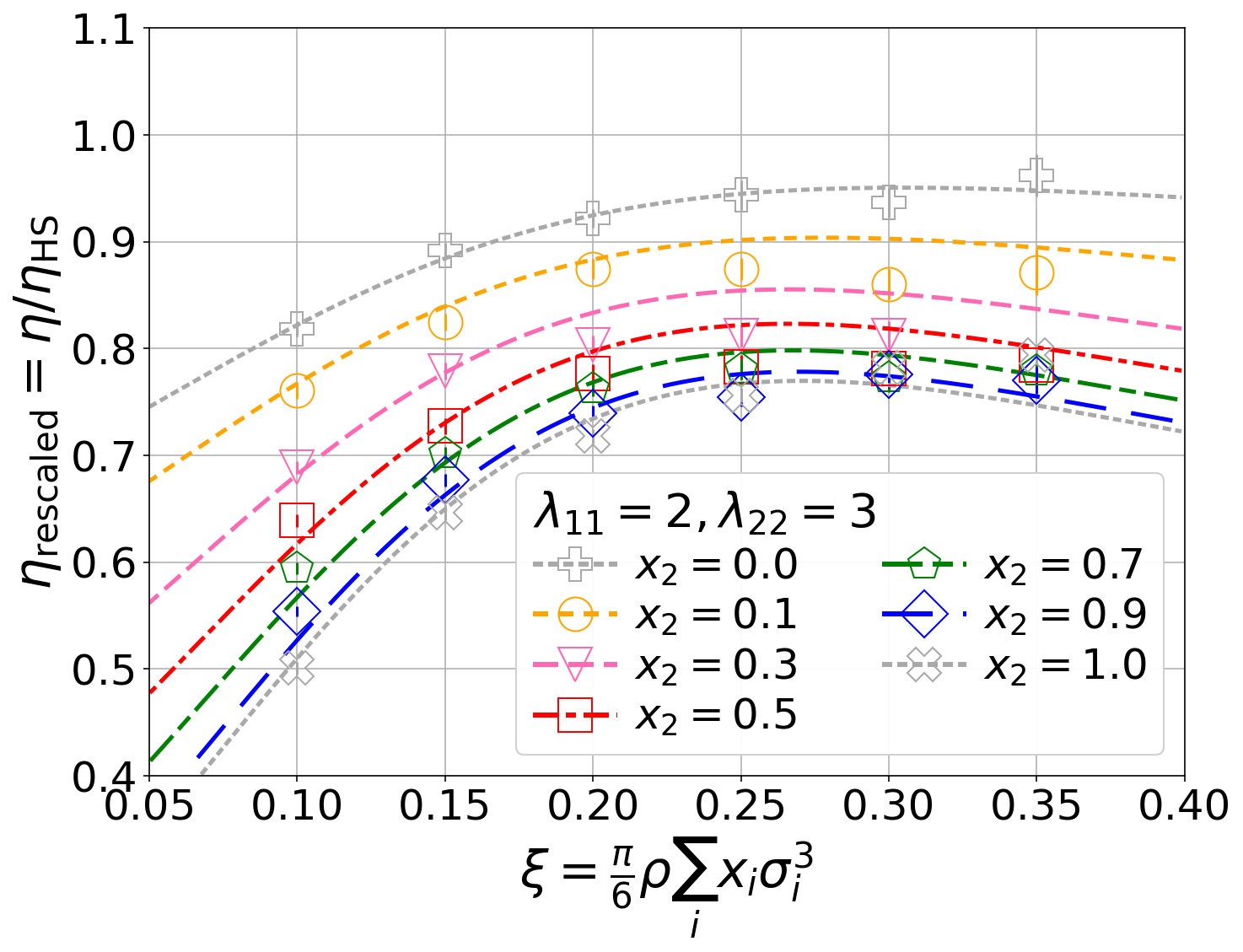}
        \label{subfig:D_different_large}}
        \newline
    \subfloat[]{
        \includegraphics[width=0.49\textwidth]{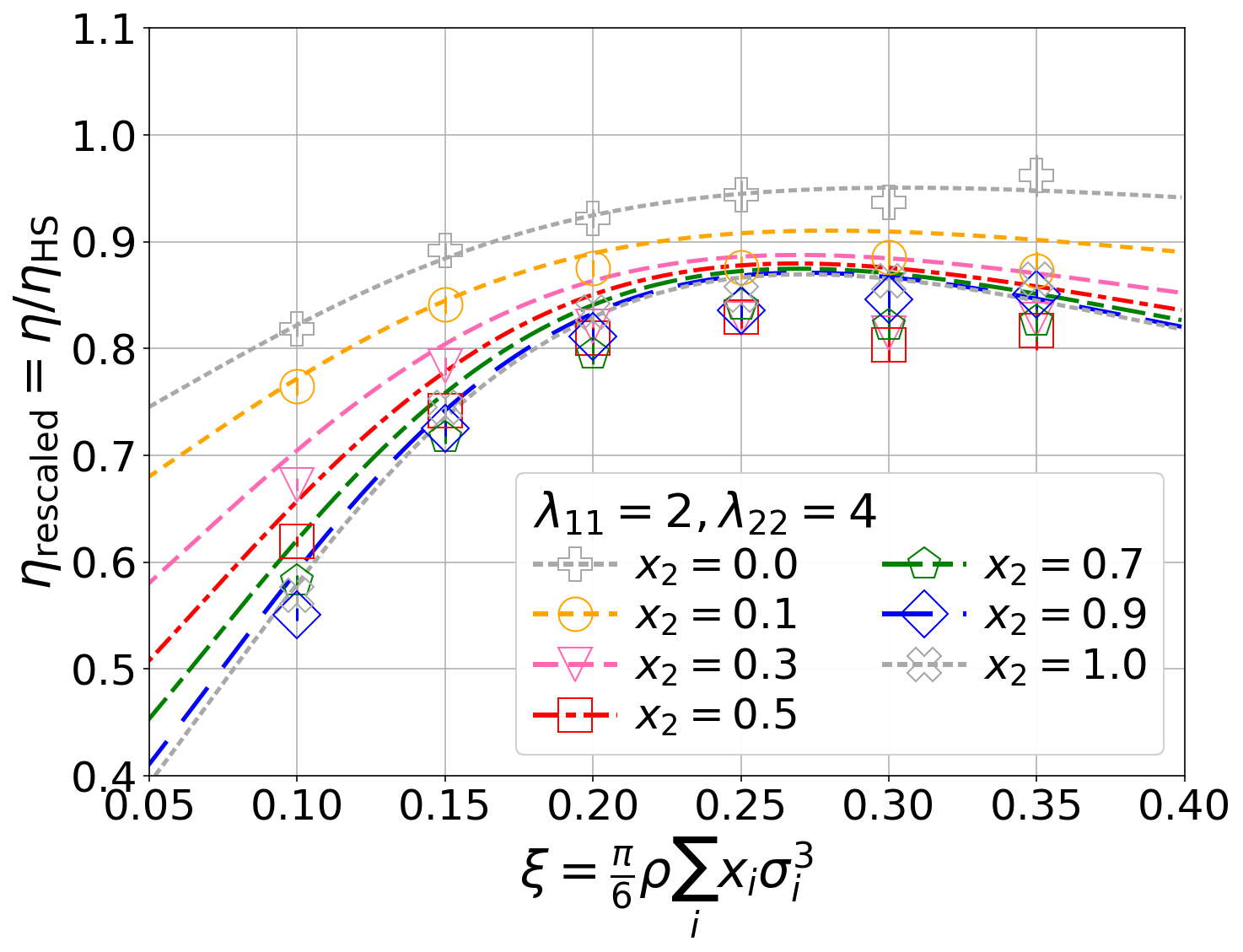}
        \label{subfig:E_different_large}}
    \subfloat[]{
        \includegraphics[width=0.49\textwidth]{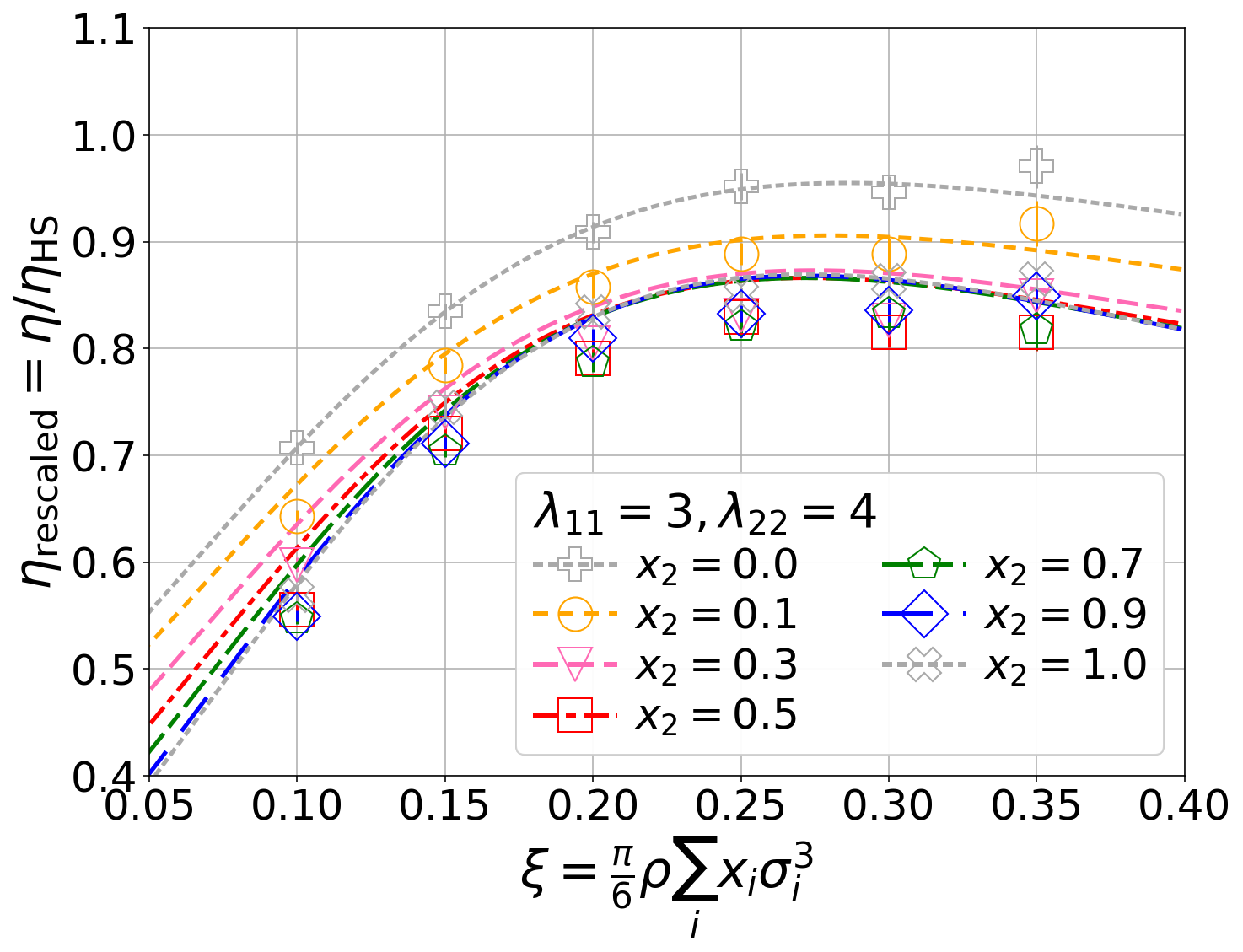}
        \label{subfig:F_different_large}}
    \caption{Rescaled shear viscosity from simulations of mixtures where the components have differently sized cores and (a) $(\lambda_{11}, \lambda_{22}) = (1, 2)$, (b) $(\lambda_{11}, \lambda_{22}) = (1, 3)$, (c) $(\lambda_{11}, \lambda_{22}) = (1, 4)$, (d) $(\lambda_{11}, \lambda_{22}) = (2, 3)$, (e) $(\lambda_{11}, \lambda_{22}) = (2, 4)$, and (f) $(\lambda_{11}, \lambda_{22}) = (3, 4)$, along with the predicted shear viscosity of the binary DHS fluid mixtures using bmDHS theory. The rescaling uses the Enskog expression for a pure hard-sphere fluid at the same volume fraction with $\sigma_i^* = 1$, mass $m_i^* = 1$, and reduced temperature $T^* = 1.5$.}
    \label{fig:Viscosity_DHS_theory_different_large}
\end{figure*}

Figures~\ref{fig:Viscosity_DHS_theory_different_small} and ~\ref{fig:Viscosity_DHS_theory_different_large} show the shear viscosity of the simulated binary DHS fluids for mixtures with two different sizes ($\sigma_1<\sigma_2$) where $\lambda_{11} > \lambda_{22}$ and $\lambda_{11} < \lambda_{22}$ respectively, along with the bmDHS predictions.
In both cases, the prediction remains accurate. However, the prediction errors are generally slightly larger compared to the case when the particle cores are of the same size, especially when the strongest dipole is associated with the large particle component.
This can, for example, be seen in Fig.~\ref{fig:Viscosity_DHS_theory_different_large}\subref{subfig:F_different_large}.
In that case ($(\lambda_{11}, \lambda_{22}) = (3, 4)$), there is a clear difference and the numerical data is systematically lower than the theoretical prediction.
It's worth noting that in Fig.~\ref{fig:Viscosity_DHS_theory_different_large}\subref{subfig:E_different_large} and Fig.~\ref{fig:Viscosity_DHS_theory_different_large}\subref{subfig:F_different_large}, the viscosities of some of the mixtures fall outside of the range of the viscosities of the pure fluids.  This could in principle be captured by Enskog-Thorne theory, and thus also by our theory, if the mixture collision integral could be obtained more accurately, either with a better mixing rule, or if it is treated as a fit parameter.

Overall, there is a high degree of correspondence between the shear viscosities from the numerical experiment and the prediction using bmDHS theory.
This good correspondence demonstrates that Enskog-Thorne theory, similarly to the pure limit Chapman-Enskog theory, can be extended to predicting the shear viscosity of dipolar fluids, even when the energy associated with the dipolar coupling is significantly larger than $k_\mathrm{B} T$.

\section{Conclusion}

We have derived an expression for the composition- and density-dependent shear viscosity of binary dipolar hard sphere (DHS) fluid mixtures based on Enskog-Thorne theory~\cite{Chapman1952} and equilibrium correlation functions.
This binary mixture DHS theory provides accurate predictions for the composition dependence of fluid mixtures based on the properties of the pure fluids.
This entails using effective parameters that are fit to the pure components and mixing rules, but no mixture-related fit parameters.

Our approach depends on an expression for the partial PDF at contact. Such expressions can generally be derived in terms of a polynomial cluster expansions.
However, due to issues with the convergence of the expansion and complexity of the higher-order terms~\cite{Novak2013}, too few coefficients are available, and these expressions are not accurate enough at even low intermediate densities.
In the present work we therefore make two physically motivated corrections to the polynomial expansion. 
Firstly, we replace the hard-sphere contribution with the far more accurate BMCSL expression for hard sphere mixtures. 
Secondly, we extract the residual dipolar contribution and re-sum this contribution using an exponential function. 

Enskog-Thorne theory additionally requires knowledge of the collision cross-section integral for unlike components. In the present work we have proposed an estimate based on the effective values obtained as fit parameters in the pure limits. Our approach relies on the observation that the effective cross-section integral can be given as a function of the effective dipolar coupling constant, independent of the moment of inertia of the hard-sphere core, at least for the range of dipolar coupling strengths and moments of inertia investigated in the present work.

Our approach demonstrates that Enskog-Thorne theory for the viscosity of hard-sphere mixtures can successfully be extended to include the impact of soft long-range dipolar interactions.
It's accuracy is mostly limited by the effective cross-section integral and the presence of the moment of inertia.

\FloatBarrier

\acknowledgments
This work has been supported by the Research Council of Norway through its Centers of Excellence funding scheme, project number 262644, and FRIPRO project number 275507.
The simulations were performed on resources provided by Sigma2 - the National Infrastructure for High-Performance Computing and Data Storage in Norway, project number NN9573K.
ASdW gratefully acknowledges the hospitality of the Nordic Institute for Theoretical Physics (Nordita) and the Nordita Corresponding Fellowship.

\end{document}